\documentclass[]{article} 
 
\usepackage{amsthm,amsmath,natbib,amssymb}

\usepackage{graphicx} 
\usepackage[FIGTOPCAP]{subfigure} 
\usepackage{comment} 

\newcommand{\mcal}[1]{\mathcal{#1}}
\newcommand{\bs}[1]{\boldsymbol{#1}}
\newcommand{\csheight}{58pt}
\newcommand{\cswidth}{0.3\linewidth}

\title{Bayesian Motion Estimation for Dust Aerosols}

\begin{document}

\maketitle
\author{\noindent Fabian E. Bachl, \emph{Heidelberg University}\\ Alex Lenkoski\footnote{Corresponding Author: alex@nr.no}, \emph{Norwegian Computing Center}\\Thordis L. Thorarinsdottir, \emph{Norwegian Computing Center}\\
Christoph S. Garbe, \emph{Heidelberg University}}\\

\begin{abstract}
Dust storms in the earth's major desert regions significantly influence microphysical weather processes, the CO$_2$-cycle and the global climate in general. Recent increases in the spatio-temporal resolution of remote sensing instruments have created new opportunities to understand these phenomena. However, the scale of the data collected and the inherent stochasticity of the underlying process pose significant challenges, requiring a careful combination of image processing and statistical techniques. In particular, using satellite imagery data, we develop a statistical model of atmospheric transport that relies on a latent Gaussian Markov random field (GMRF) for inference.  In doing so, we make a link between the optical flow method of Horn and Schunck and the formulation of the transport process as a latent field in a generalized linear model, which enables the use of the integrated nested Laplace approximation for inference. This framework is specified such that it satisfies the so-called integrated continuity equation, thereby intrinsically expressing the divergence of the field as a multiplicative factor covering air compressibility and satellite column projection. The importance of this step -- as well as treating the problem in a fully statistical manner -- is emphasized by a simulation study where inference based on this latent GMRF clearly reduces errors of the estimated flow field. We conclude with a study of the dynamics of dust storms formed over Saharan Africa and show that our methodology is able to accurately and coherently track the storm movement, a critical problem in this field.
\end{abstract}

\clearpage
\section{Introduction}\label{sec:intro}

Dust storms are a global meteorological phenomena originating from arid and semi-arid regions. They interfere with human modes of living and transportation, alter the radiation transmittance and circulation of the earth's atmosphere and interact with microphysical cloud processes. Moreover, dust deposition provides vital nutrients for microorganisms that ultimately influence the CO$_2$-cycle. The detection of dust storms, the prediction of their development, and the estimation of sources  are therefore of immediate interest for a wide range of environmental applications. Remote sensing systems play an indispensable role in characterizing the dynamics of these systems, thereby providing the raw data that enables statistical analysis. This article discusses the development of a Bayesian hierarchical framework that uses remote sensing data to detect dust plumes, track their movement and pinpoint their source.  

\begin{figure*}[]
  \centering
  \subfigure[]{
    \label{PlumeJan18super}
    \includegraphics[width=0.8\linewidth]{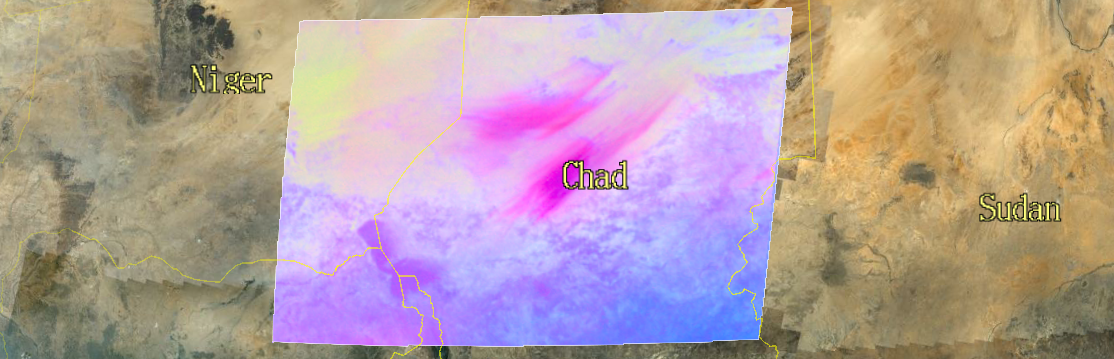} 
  } 
  
  \subfigure[]{
    \label{PlumeJan_18_0730GMT}
    \includegraphics[width=0.25\linewidth]{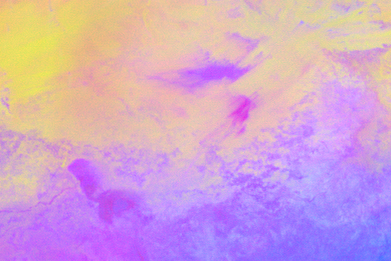} 
  }
  \subfigure[]{
    \label{PlumeJan18_0830GMT}
    \includegraphics[width=0.25\linewidth]{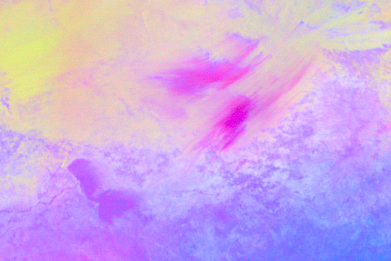} 
  }
    \subfigure[]{
    \label{PlumeJan18_0930GMT}
    \includegraphics[width=0.25\linewidth]{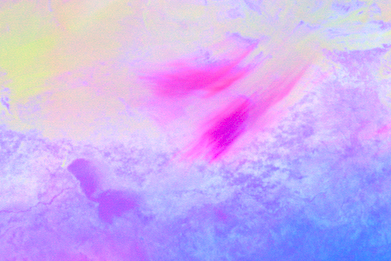} 
  }
  \caption{(a) SEVIRI falsecolor imagery according to \cite{Lensky2008Clouds} superimposed on a Google Earth depiction of Niger, Chad and Sudan. The central pink area is a dust plume emerging on January 18, 2010 at about 9.30 am GMT over Chad.  Panels (b) to (d) visualize the development of the plume at 7.30 am, 8.30 am and 9.30 am GMT, respectively.}
  \label{fig:PlumeJan182010}
\end{figure*}

In the case of dust aerosols, the Meteosat series of satellites and, in particular, the Spinning Enhanced Visible and InfraRed Imager (SEVIRI) aboard the geostationary Meteosat-9 poses an unique opportunity as it is the first time that the respective spatial and temporal coverage allows for the analysis of local and sub-daily processes of dust emission and transport. Alongside visible spectra, SEVIRI provides infrared measurements at frequencies from 3.9 to 13.4 $\mu$m every 15 minutes at a spatial resolution of 3 km at nadir. Figure \ref{fig:PlumeJan182010} shows a visual depiction of the so called SEVIRI falsecolor imagery (SFI), a common mode of visually assessing dust aerosols, which form the basis of our data. 

The contemporary analysis of dust aerosols follows two different paradigms. Motivated by physical models of conditions for dust emission, transport via wind fields and radiative filtering properties of aerosols, the work of \cite{Klueser2009Remote},  and \cite{Brindley2012critical} is based on connections between SFI and aerosol optical depth (AOD). Here, the presence of dust is quantified by a combination of different SFI thresholds derived from case- and simulation-studies. In contrast, the work of  \cite{Rivas-Perea2010Traditional} and \cite{Eissa2012Dust} employs methods from machine learning and image processing by using neural nets to learn non-linear dust detection criteria from a data set with labels set by a human expert. 

From a statistical viewpoint, both approaches suffer shortcomings. Directly imposing thresholds partly based on expert opinion might lead to misleading conclusions due to human subjectivity. Also, neither \cite{Klueser2009Remote} nor \cite{Brindley2012critical} include quantification of uncertainty in their analysis. Neural nets, on the other hand, are directly driven by data and interpretable in a probabilistic sense.  However, these methods are often criticized for a lack of transparency and non-physical motivation which in turn obfuscates scientific interpretability.

Further, none of the previously mentioned approaches imposes a coherent spatio-temporal structure. As a respective smoothness assumption can easily be justified by the corresponding transport process, this omits valuable information. Previous attempts, e.g. by \cite{Schepanski2007new}, to localize and characterize areas being sources of dust storms have to rely on human visual data inspection and indication. \cite{Bachl2012Classify} show that a sufficiently accurate data driven estimation of the dust flux allows automation of this process and can quantify dust source output. Dust flux may also be employed to perform hazard forecasts, to interpolate areas with missing observation data (such as areas covered by clouds), or to validate atmospheric wind field based models. 

Various approaches in different scientific fields capture similar problems and are closely related to our methodology. Statistical approaches are predominantly driven by applications related to either the verification of numerical weather predictions or the issuing of so-called nowcasts, forecasts for very short lead-times, see e.g. \cite{Gilleland2010Analyzing} and \cite{Xu2005Kernel-Based}. Here, a transformation between two spatial fields (e.g. a prediction and the corresponding observation) is determined via a deformation field that associates spatial locations of the two fields in a smooth fashion. In prediction problems, the deformation field then serves as a tool to assess the field both in terms of mis-localization and quantification error. In contrast, in nowcasting a current spatial observation and a given deformation field are utilized to predict the spatial field representing future realizations. 

\cite{Xu2005Kernel-Based} apply a integro-difference equation where information is propagated between the two fields through a kernel function. In image processing differential approaches--which can be interpreted as special cases of the integro-difference equation--have been popular since the advent of the Optical Flow (OF) method of \cite{Horn1981Determining}, for brevity called HS-OF from here on.  These methods have already entered the statistics community, see e.g. \cite{Marzban2010Optical} who employ the connate OF approach of \cite{Lucas1981iterative}.

In this contribution we begin with the HS-OF method and illustrate how to formulate this approach as a Bayesian hierarchical model.  This gives a probabilistic interpretation of the optical flow as a latent Gaussian Markov random field. While the link is relatively straightforward, to the best of our knowledge, this is the first time that the full distributional aspects and the associated uncertainty are taken into account for the HS-OF method. The intrinsic smoothness parameter of the method then finds a clear meaning as the precision hyper-parameter of the conditional autoregression model imposed on the flow. This perspective comes with several long and short term benefits. Firstly, inference can be performed using computationally efficient integrated nested Laplace approximations (INLA) \citep{Rue2009Approximate}. A second benefit is the interpretability of the flow field in terms of the physical nature of the phenomenon under consideration.

Our second contribution is to leverage the hierarchical Bayesian framework to overcome deficiencies in the HS-OF formulation.  A typical quirk of statistical warping and optical flow is the underlying preservation assumption of the respective quantity along its trajectory. In dust aerosols (as well as other natural phenomenon) this might lead to false conclusions. Gaseous solutions are compressible and remote sensing often only leads to a non-bijective mapping of a three dimensional quantity to a two dimensional data space. Alongside advection, observations are therefore clearly prone to convective effects resulting from compression of the solution or material exchange inside a projected atmospheric column. As a remedy we extend the HS-OF method to incorporate the water vapor related work of \cite{Corpetti2002Dense} and put it in a Bayesian hierarchical model context. As our work emphasizes by a simulation study, the Integrated Continuity Equation (ICE) considerably reduces errors in the estimated flow field. The main advantage of the ICE comes from the fact that it implicitly considers a multiplicative convective effect that is driven by the divergence of the flow field itself. A motion trajectory starting at a point where the divergence is positive (negative) leads to a low (high) multiplicative effect mimicking the dispersion (accumulation) of the modeled quantity. In case of pure advection, i.e. the absence of divergence, the multiplicative factor is 1 and the usual preservation assumption is retained.

The article proceeds as follows. Section 2 offers a description of the data. The following Section 3 is two-fold. As a first step it illustrates the basic thresholding concept for dust detection as well as our approach to employ a generalized linear model for this task. The Horn and Schunck method for motion estimation is then reviewed and extended by the concept of the integrated continuity equation and a probabilistic interpretation of both approaches is provided. In Section 4, we evaluate our framework in three ways.  First, we assess the detection method in comparison to thresholding and linear discriminant approaches. Section 4 then focuses on a simulation study analyzing Bayesian inference of the motion estimation techniques mentioned above. Finally, we show results of applying ICE motion estimation to dust detected from SEVIRI measurements. The last Section 5 provides a discussion of our results and future work.
\section{Data and Operative Products}\label{sec:data}
The SEVIRI instrument resides aboard the Meteosat-9 satellite launched on December 21, 2005 in a joint effort of the European Organization for the Exploitation of Meteorological Satellites (EUMETSAT) and the European Space Agency (ESA). Being an integral part of the payload of the Meteosat Second Generation (MSG) series of platforms dedicated to environmental data collection, SEVIRI measures electromagnetic radiation at 12 different spectral windows spanning from visible to infrared frequencies \citep{Schmetz2002Introduction}. With Meteosat-9 residing at 0 degrees of latitude, 0 degrees of longitude and a height of approximately 36 km it provides measurements for up to approximately 80 degrees of deviation from nadir where it has a resolution of about $3 \times 3$ km. In combination with the per-image scan time of 12 minutes and three minutes of calibration this results in a $3712 \times 3712$ pixel imagery every 15 minutes.

With respect to radiative remote sensing, the most dominant effect of dust aerosols is to filter the infrared radiation leaving the terrestrial surface in a frequency dependent fashion. This phenomenon is reflected by the 12.0 $\mu$m, 10.8 $\mu$m and 8.7 $\mu$m channels of the SEVIRI instrument, which we will abbreviate by $BT_{12.0}$, $BT_{10.8}$ and $BT_{8.7}$, respectively. For example, it is well known that in the presence of dust aerosols the difference $\Delta T_{BR} = BT_{12.0}-BT_{10.8}$ increases while $\Delta T_{BG} = BT_{10.8}-BT_{8.7}$ decreases \citep{Schepanski2007new}. This connection, also known as split window technique, results in popular operative products such as the SFI (see Figure \ref{fig:PlumeJan182010}) which defines the red ($R$), green ($G$) and blue ($B$) channels of the visualization as
\begin{eqnarray*}
R &=& L_R(\Delta T_{BR}),\\
G &=& L_G(\Delta T_{BG})^{\gamma},\\
B &=& L_B(BT_{10.8}),
\end{eqnarray*}
where $L_c$ with $c \in \{R,G,B\}$ are linear rescaling functions (see \cite{Lensky2008Clouds} for further details) and $\gamma=0.4$.

As indicated by a simulation study performed by \cite{Brindley2012critical}, this leads to a correlation between the tendency of the SFI to appear pink and the optical depth $\tau_{10}$ of the atmosphere at 10 $\mu$m being increased by the presence of dust aerosols. The same study also analyzes how daytime, surface vegetation, seasonal atmospheric conditions, satellite viewing angle and plume height enter the SFI product.  

Recently, \cite{Ashpole2012automated} proposed an extended thresholding scheme for dust detection given by 
\begin{eqnarray}
\Delta T_{BR}&>&0 K, \label{eq:ashpoleThresh1}\\
\Delta T_{BG}&<&10 K,  \label{eq:ashpoleThresh2}\\
BT_{10.8}&<&285 K,  \label{eq:ashpoleThresh3}\\
\Delta T_{BR}-M&<&-2 K,  \label{eq:ashpoleThresh4}
\end{eqnarray}
where $K$ denotes the unit of brightness temperature in Kelvin.  Alongside requiring the fixed conditions given in Equations \eqref{eq:ashpoleThresh1} and \eqref{eq:ashpoleThresh2} in order to flag a pixel to contain dust they introduce two additional requirements. Since the blue channel is generally saturated in the presence of dust while the occurrence of clouds lowers its brightness, the threshold $BT_{10.8}<285 K$ in Equation \eqref{eq:ashpoleThresh3} removes artifacts coming from the latter. The last threshold is data dependent and serves two purposes. By requiring Equation \eqref{eq:ashpoleThresh4} to hold, where $M$ is a two-week cloud masked rolling mean of $\Delta T_{BR}$, it rules out false positive dust detections where clouds are present and over regions where the red channel is close to saturation even under pristine conditions.


\section{Methods}\label{sec:methods}

This section is two fold. Firstly, we go into detail about dust aerosol detection, i.e. the task of assigning a given pixel of the SEVIRI imagery with a quantity representing the evidence of the presence of dust aerosols. The second part then elaborates how this quantity may be employed to infer the motion of the aerosol by modeling the underlying transport process relying on a differential perspective.

\subsection{Dust detection using generalized linear models}\label{sec:dust_detection}

Let $\mcal{S} \subset \mathbb{R}^2$ denote the image domain and assume we have a series of images obtained over the time interval $[0,T]$.  Our first goal is to determine the dust indicator variable $d_{xyt}$ with $d_{xyt}=1$ if location $(x,y)\in\mcal{S}$ is covered by a dust plume at time $t \in [0,T]$ and $d_{xyt}=0$ otherwise.  This assessment is made on the basis of the observation vector $\bs{I}_{xyt} = (I_{1xyt}, I_{2xyt}, I_{3xyt})$ where the three components of $\bs{I}_{xyt}$ correspond to the red, blue and green channels as discussed in Section~\ref{sec:data}. Since the surface in $\mcal{S}$ is naturally varied, a critical component in determining $d_{xyt}$ is the background appearance $\bs{A}_{xyt}$ at each location $(x,y)\in\mcal{S}$ and time $t$ when no dust or cloud cover exists.  The background is compared to $\bs{I}_{xyt}$ to assess whether a dust plume covers the location at time $t$.

Our method of detecting dust aerosols is a progressive refinement of linear discriminant analysis (LDA), which infers projection coefficients $r_i$ and an offset $q$ such that the sign of 
\begin{equation}
\qquad \eta(x,y,t) = q + \sum_{i = 1}^3 I_{ixyt} \,  r_i 
\label{eq:LDA}
\end{equation}
serves as a label for the dust content of a particular location. In \cite{Bachl2012Classify}, a three level Bayesian hierarchical model is developed where the projection coefficients and intercepts are functions of the appearance estimate $\bs{A}_{xyt}$.  In the first level, the dust indicator variable $d_{xyt}$ is modeled by
\[
\mathbb{P}(d_{xyt}=1) = \mathrm{logsig}(\eta(x,y,t)), 
\]
where $\mathrm{logsig}$ denotes the log-sigmoid transfer function and 
\begin{equation}
\qquad \eta(x,y,t) =\sum_{i = 1}^3 I_{ixyt} \,  f_i^1(A_{ixyt}) + f_i^2(A_{ixyt}).
\label{eq:ESA2012LP}
\end{equation}

The second and third levels are prior distributions on the latent functions $f_i^1$ and $f_i^2$ and their parameters, respectively. The functions $f_i^1$ and $f_i^2$ are modeled semi-parametrically by binning each component of $\bs{A}_{xyt}$ into 100 distinct bins taken over the range of each component over the image $\mcal{S}$.  These functionals are then modeled as continuous random walks (CRWs), that is,
\begin{eqnarray*}
f_i^1& \sim & \mathcal{N}_{100}(0,Q_{\mathrm{CRW}}(\mathbf{\Theta}_i)), \\
f_i^2 & \sim  & \mathcal{N}_{100}(0,Q_{\mathrm{CRW}}(\mathbf{\Xi}_i)). 
\end{eqnarray*}
The parameters of the CRWs, $\mathbf{\Theta}_i$ and $\mathbf{\Xi}_i$, are given independent log gamma priors. 

\cite{Bachl2012Bayesian} note that a drawback of this approach is that the signal noise in (\ref{eq:ESA2012LP}) is carried over in a linear fashion which can hamper consecutive motion estimation. As a remedy they propose to shift the SFI as to be a part of the domain of the latent functions such that 
\begin{equation}
\eta(x,y,t) = \sum_{i=1}^3 h_i(A_{ixyt},A_{ixyt} - I_{ixyt}),
\label{eq:IGARSS2012LP}
\end{equation}
where the domain of $A_{ixyt} - I_{ixyt}$ is discretized in a manner similar to that of $A_{ixyt}$. The functions $h_{i}$ are modeled as a two dimensional conditional autoregression (CAR) intrinsic GMRFs (see \cite{Rue2005Gaussian} for details) with
\[
p(h_{i}(j,k)) \propto \exp\Big(-\rho\sum_{(l,m)\sim (j,k)} \big(h_{i}(l,m)-h_{i}(j,k)\big)^2\Big),
\]
where ``$\sim$'' denotes the four nearest neighbors on the two dimensional discretization grid of $A_{ixyt} \times (A_{ixyt} - I_{ixyt})$. 

Yet, as discussed in \cite{Bachl2013Bayesian}, the estimation of the background radiation remains a critical aspect. Alongside the cyclic issue of requiring a criterion to mark a region as dust free, the radiative characteristics of this region generally vary even under pristine conditions. However, the vegetative properties of the largely unpopulated African continent significantly determines the general appearance of the SFI.  The study therefore proposes to employ the monthly average surface emissivity $\bs{E}_{xyt}$ (see Figure \ref{sammplesEmissPristine}) product at $8.4 \mu$m according to \cite{Seemann2008Development}, which strongly correlates with the vegetation, to supersede the anomaly indicating term $A-I$, hence
\begin{eqnarray}
\eta(x,y,t) &=& \sum_{i=1}^3  g_i(I_{ixyt},E_{ixyt}),
\label{eq:IGARSS2013LP}
\end{eqnarray}
where the new functional $g_i$ is modeled in a manner similar to $h_i$ above. In the following, we refer to the model in \eqref{eq:IGARSS2013LP} as the latent signal mapping (LSM) approach. 

In practice, we are therefore required to determine several quantities, namely the background appearance $\bs{A}_{xyt}$ or the emissivity $\bs{E}_{xyt}$, and subsequently fit a statistical model for $\eta(x,y,t)$ using training data.  Estimating the latter is performed by using a large set of labeled training data, see Figure~\ref{fig:labeledData} for an example of one image used in our training set.

\subsection{Motion Estimation}

Rheology, the study of the flow of liquid matter, and the motion estimation of quasi-rigid bodies has been a very active research field of image processing and computer vision during the last two decades. With respect to image analysis in experimental fluid dynamics these efforts led to an increasing expertise in correlation-based particle image velocimetry methods and variational approaches to the problem. See \cite{Heitz2010Variational} for a review on this topic. A similar effect has occurred to computational statistics due to the increasing interest in modeling spatio-temporal processes for environmental science applications, e.g. ozone and precipitation interpolation and forecasting. In particular, methods based on the perspective of warping have constantly been developed further, see e.g. the review by \cite{Glasbey1998review} and the work of \cite{Aberg2005image}. 

However, to the best of our knowledge, the connection between probabilistic and variational approaches is reflected only by a few publications.  \cite{Simoncelli1991Probability} point out the distributional aspects of the well-known Horn and Schunck (HS) method of optical flow \citep{Horn1981Determining}. A maximum-posteriori approach to the free parameters of this method was illustrated by \cite{Krajsek2006maximum} through the use of a Bayesian hierarchical model. \cite{Krajsek2006equivalence} further show the limit-equivalence of the variational solution of the HS functional to the mode of a normal distribution defined via the maximum entropy principle with respect to observations at discretized locations. 

\subsubsection{The Horn and Schunck Approach to Optical Flow} Once the linear predictors of dust probability $\eta(x,y,t)$ are determined, it is helpful to model their dynamics in both space and time.  This allows the projection of dust storm probabilities to $\eta(x',y',\tau)$ for all $(x',y')\in\mcal{S}$ and $\tau \in [0,T]$ allowing one to project dust probabilities forward as well as ``rewind'' the storm to determine its source.

As above, fix $(x,y)\in\mcal{S}$.  We then aim to determine the vector field $\bs{w}(x,y,t) = (u(x,y,t), v(x,y,t))$, where $u(x,y,t)$ and $v(x,y,t)$ are the instantaneous change in $\eta(x,y,t)$ in the vertical and horizontal directions.  As discussed in Section~\ref{sec:intro}, we follow the motion estimation literature in our development and subsequently show that it is related to the Bayesian estimation of spatially dependent random effect models.


Most motion estimation techniques are based on the assumption that there is a photometric or geometric quantity of the image sequence that is preserved spatially or temporally. In case of HS optical flow this is expressed by the brightness constancy equation (BCE).   For a given triplet $(x,y,t)$, suppose that $\eta(x,y,t) = k$.  The BCE formulation then stipulates that there is a path in $\mcal{S}$, (x(r), y(r)) for all $r \in [0,T]$ such that
\begin{equation}
\eta(x(r),y(r),r) = k.
\label{eq:BCE}
\end{equation}
Thus, the total derivative of the intensity function with respect to time vanishes. Assuming no higher order dependencies of $x$ and $y$ (i.e. $\mathrm{d}x/\mathrm{d}t = \partial x/ \partial t $ and $\mathrm{d}y/\mathrm{d}t = \partial y/ \partial t $) it holds that
\begin{eqnarray*}
0 = \frac{\mathrm{d}}{\mathrm{d}t} \eta &=& \frac{\partial}{\partial t} \eta + \frac{\mathrm{d} x}{\mathrm{d}t} \frac{\partial}{\partial x} \eta + \frac{\mathrm{d} y}{\mathrm{d}t} \frac{\partial}{\partial y} \eta \\
&=& \eta_t + \frac{\mathrm{d} x}{\mathrm{d}t} \eta_x + \frac{\mathrm{d} y}{\mathrm{d}t} \eta_y \\
&\approx& \eta_t + u \, \eta_x + v \, \eta_y,
\end{eqnarray*}
where the dependence on $(x,y,t)$ has been dropped.

This equation is, however, under-determined, an issue known as the aperture problem. As with many other approaches, the HS optical flow therefore imposes an additional constraint. In order to maintain physical plausibility and to propagate information into image regions with ambiguous gradient properties, non-smoothness of the flow is penalized via the Euclidean norm of the gradient. The final optical flow is then defined as the minimizer of the weighted average squared deviations of the BCE fit integrated over the image domain $\mcal{S}$. That is, 
\[
(u,v)(\alpha) = \mathrm{argmin}_{u,v} L_\mathrm{HS}(\alpha),
\]
where $\alpha$ is a regularization parameter and 
\[
L_\mathrm{HS}(\alpha) = \int_{\mcal{S}} (\eta_t + u \, \eta_x+v  \, \eta_y)^2 + \alpha^2 (|\nabla u|^2 +|\nabla v|^2).
\]
Existence and uniqueness of the minimizer were shown by \cite{Schnorr1991Determining} under mild restrictions on $\eta$ and $(u,v)$ in terms of Sobolev spaces. 

In the discrete sense the BCE error term is equivalent to an interpretation of the image gradients as an observational system of the latent flow variables $u$ and $v$ with additive Gaussian noise, that is,
\[
u \, \eta_x + v \, \eta_y = -\eta_t + \epsilon, \qquad \epsilon \sim\mathcal{N}(0,\sigma^2).
\]
It follows that the partial derivatives of $\eta(x,y,t)$ define a Gaussian likelihood $p(\nabla \eta|\mathbf{u},\mathbf{v})$ for the discretized optical flow. Focusing on the regularization term and using forward differences as discrete lattice approximations to the local flow gradients $\nabla u$ and $\nabla v$, the directional components of the functional $L_\mathrm{HS}$ reduce to
\[
\int_{\mcal{S}} \alpha^2 |\nabla u|^2 \approx \alpha^2 \sum_{s\in\mcal{S}} (u_s-u_{n(s)})^2,
\]
where $n(s)$ is the corresponding neighbor on the discretization grid, and similar for $v$. This formulation is analytically identical to the log-density of a CAR GMRF, illustrating the equivalence of the estimation of HS optical flow estimation and Bayesian modeling of spatially dependent systems \citep{Besag1974Spatial}. Thus, the smoothness part of the HS functional defines intrinsic GMRF priors $p(u) = \mathcal{N}(\mathbf{0},Q_u)$ and $p(v) = \mathcal{N}(\mathbf{0},Q_v)$ for the latent flow fields if the precision matrices are defined via
\begin{eqnarray*}
Q_{ij}(\alpha) = \alpha^2 \begin{cases} n_i, & i=j  \\ -1, &  i \sim j \\ 0, & \mathrm{otherwise} \end{cases}
\end{eqnarray*}
where $n_i$ is the number of neighbors on the grid. 
This formulation also clarifies the role of the smoothness parameter $\alpha$ as a hyper parameter of the precision matrix $Q$. Assuming independence from other variables of the model, the optical flow is thus given as the posterior
\[
p(\mathbf{u},\mathbf{v} | \nabla \eta) \propto \int p(\nabla \eta  |  \mathbf{u},\mathbf{v})p(\mathbf{u},\mathbf{v} |  \alpha)p(\alpha) \, \textup{d}\alpha.
\]
%
%

\subsubsection{The Integrated Continuity Model}
\label{sec:ICE}
While HS optical flow, and particularly the BCE assumption, is sufficient to model motion of rigid bodies in many areas of image processing, it is clearly insufficient in capturing the dynamics of $\eta(x,y,t)$.  Constancy of image brightness implies that the flux of the quantity under consideration is divergence-free. This assumption is often violated for two reasons. On the one hand, the observed material itself might be compressible, as is the case for dust aerosols. Alternatively, even if incompressible fluids like water are considered, the imaging technique might deliver a two-dimensional projection of a three-dimensional process. Thus, even if this process obeys a divergence-free flow, the projection might miss strong sources and sinks due to the fluid convection through the layers of the $z$-axis. 

The general idea of the integrated continuity equation (ICE) \citep{Corpetti2002Dense} is that it relates the local intensity change to the flux of the quantity through the boundary surface of an infinitesimal volume. That is, it assumes
\begin{eqnarray*}
0 = \frac{\mathrm{d}}{\mathrm{d} t} \eta &=& \eta_t + \mathrm{div}(\eta \mathbf{w})\\
&=& \eta_t + \eta u_x + \eta_x u + \eta v_y + \eta_y v \\
&=& [\mathbf{w},1] \cdot \nabla \eta + \eta \, \mathrm{div}(\mathbf{w}). 
\end{eqnarray*}
This equation also shows the connection to the BCE as it reduces to the former for incompressible materials when the divergence of $\mathbf{w}$ fulfills $\mathrm{div}(\mathbf{w})=0$. 

Following \cite{Corpetti2002Dense} the flow according to the ICE is then defined as the minimizer of the functional 
\[
L_\mathrm{ICE}(\alpha) = \int_\mcal{S} ([\mathbf{w},1] \cdot \nabla \eta + \eta \, \mathrm{div}(\mathbf{w}))^2 + \alpha^2 (|\nabla u|^2 +|\nabla v|^2).
\]
Using the discrete divergence approximation
\[
\text{div}([u,v]_{ij}) \approx \frac{1}{2} \big((u_{i,j+1}-u_{i,j-1}) + (v_{i+1,j}-v_{i-1,j})\big)
\]
leads to the following likelihood equation of the flow field given the image
\[
u_{ij} \eta_x+ v_{ij} \eta_y + \frac{\eta}{2}\big((u_{i,j+1}-u_{i,j-1}) + (v_{i+1,j}-v_{i-1,j})\big) = -\eta_t + \epsilon_{ij},
\]
where again $\epsilon_{ij} \sim \mathcal{N}(0,\sigma^2)$.

In what follows we show that the ICE approach to determining optical flow of dust storms considerably improves estimated flow fields obtained using HS methods, largely for the obvious reasons that dust storms grow and then diminish through time.  As should be clear from the development, estimation of the posterior distribution $p(\mathbf{u},\mathbf{v}|\nabla \eta)$ for the flow vector fields under either the HS or ICE paradigms is easily performed using the INLA methodology \citep{Rue2009Approximate}.

\section{Applications}\label{sec:apps}
We now proceed with a series of studies that investigate the performance of the individual components of our framework and conclude with a set of case studies that show how the entire system performs at detecting and tracking dust storms.  Section~\ref{sec:apps_detect} focuses on the storm detection component--the model for determining $\eta(x,y,t)$--and compares our method with several reference methods.  Section~\ref{sec:apps_flow} then conducts a simulation study (since ground truth of vector fields is unavailable) that assesses the performance of the ICE formulation of optical flow over the original HS formulation.  Finally, we conclude in Section~\ref{sec:apps_case} with an in-depth investigation of two dust storms and show how our method is able to correctly identify the storm, and model its flow.
\subsection{Aerosol Detection}\label{sec:apps_detect}

\begin{figure}[h]
\center
    \includegraphics[width=0.9\linewidth]{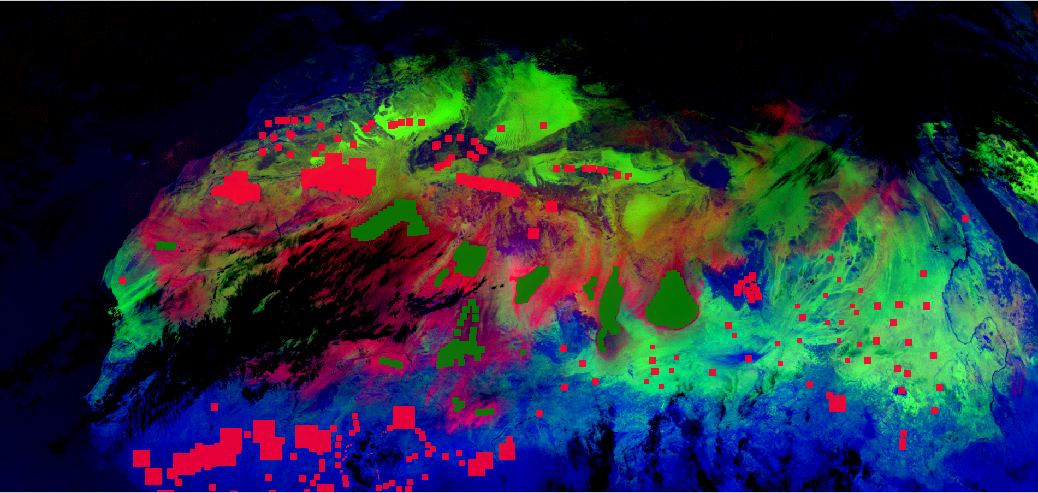}
    \caption{Training data in falsecolor representation. Red: pixels labeled as pristine. Green: pixels labeled as dusty.}
  \label{fig:labeledData}
\end{figure}

The basis of the following analysis is a SEVIRI data set spanning January 10--26, 2010, a period with several smaller and large scale dust events. By visual inspection we performed an extensive labeling of dusty and pristine regions. As the SEVIRI signal changes strongly with the relative position of the sun and dust plume genesis often predominantly occurs during the forenoon, a corresponding subsampling of pixels according to the time stamps of the complete satellite imagery and the local latitude was performed. An example for a labeled frame of the sequence is given in Figure \ref{fig:labeledData}. 

\begin{figure*}[!bhpt]
  \centering
  \subfigure[Dusty pixels]{
    \label{PlumeJan18}
    \includegraphics[width=0.4\linewidth]{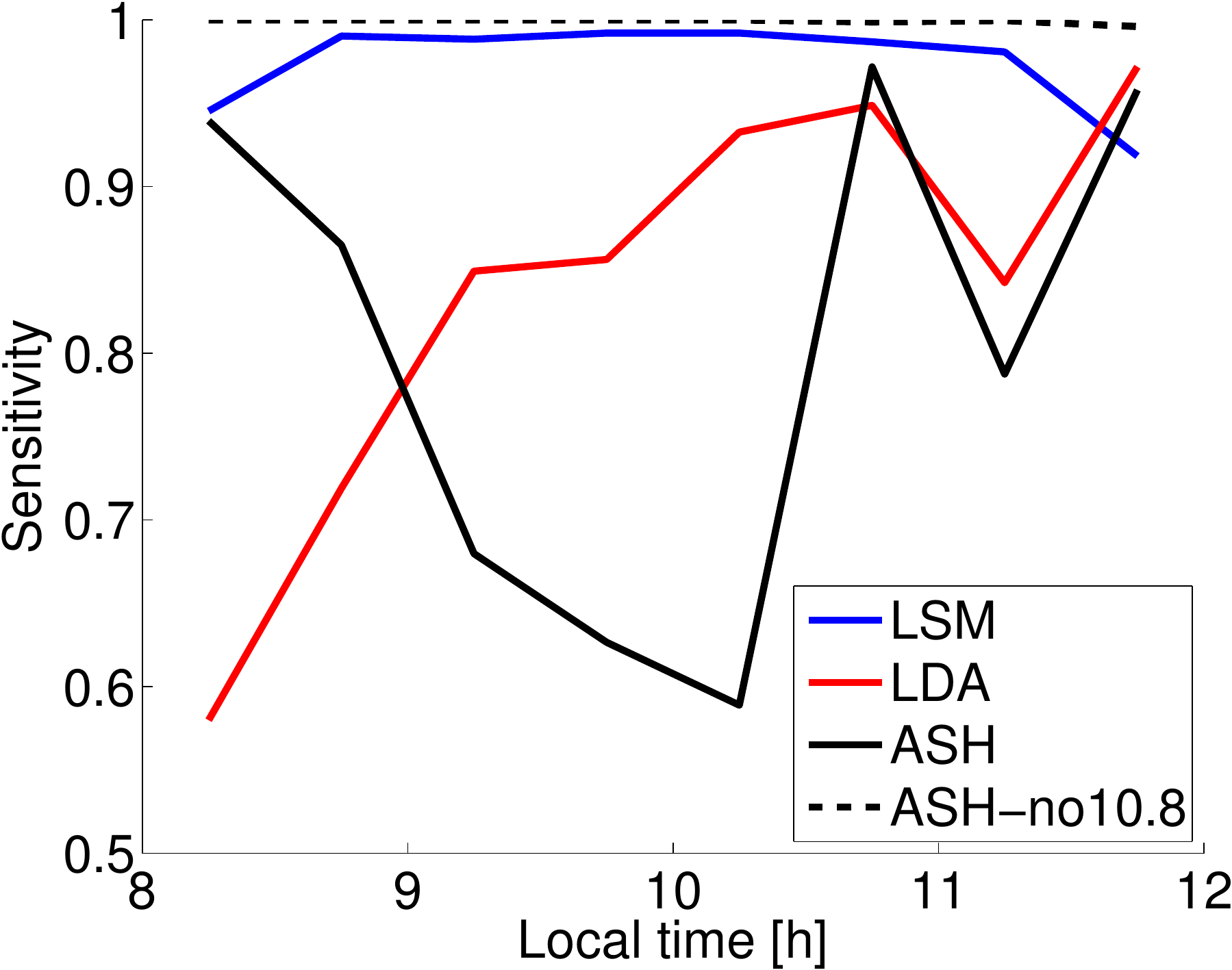} 
  } 
  \subfigure[Pristine pixels]{
    \label{PlumeJan18}
    \includegraphics[width=0.4\linewidth]{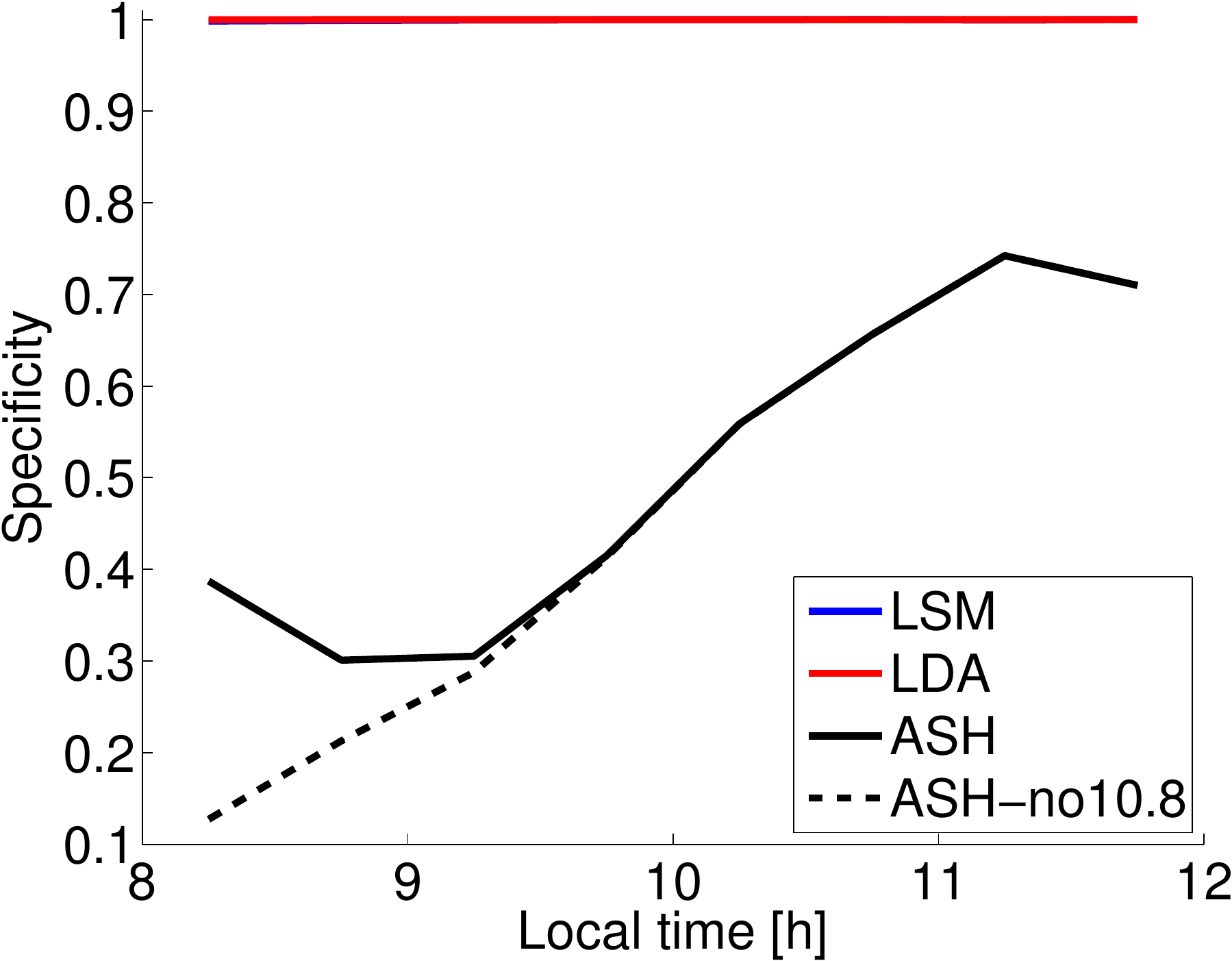} 
  }    
  \caption{Cross validation results for pixel-wise dust detection under the LSM emissivity approach (blue), linear discriminant analysis (red) and the two thresholding methods of \cite{Ashpole2012automated} (black). The plots show the percentage of correctly classified (a) dusty pixels and (b) pristine pixels, stratified by the hour of the day.}
  \label{testData}
\end{figure*}

After labeling these images, we conducted a two-fold cross validation study.  In each case, we trained the models on a subset of the data and then judged the fit on those observations left out.  We compared the performance of four methods for estimating the probability of dust, the latent signal mapping (LSM) approach of (\ref{eq:IGARSS2013LP}), a simple linear discriminant analysis (LDA), and two thresholding approaches introduced by \citet{Ashpole2012automated}. In case of LDA and LSM a pixel is classified as dusty if the probability of dust is greater than 0.5, and as pristine otherwise. The first approach of Ashpole and Washington (ASH-no10.8) determines a pixel to be dusty if Equations \ref{eq:ashpoleThresh1} and \ref{eq:ashpoleThresh2} hold. For the second method (ASH), also \ref{eq:ashpoleThresh3} is required to hold. Figure~\ref{testData} shows the percentage of correctly classified clear pixels (left panel) and those containing dust (right panel), stratified by the time of day of the image.  


From Figure~\ref{testData} we draw several interesting conclusions.  First, we see that the two thresholding approaches perform poorly in correctly classifying clear, or pristine, regions. Even the more involved ``ASH'' leads only to slight improvements. By contrast, the simpler ``ASH-no10.8'' thresholding approach performs essentially perfectly at classifying clear regions while the additional threshold of ``ASH'' significantly decreases the fraction of correctly recognized dusty samples. By contrast, the LDA perfectly classifies pristine areas, but performs poorly during the early hours (between 8 am and 10 am) at classifying dusty pixels.  Finally, the LSM method considerably improves on LDA for dusty pixels and achieves nearly perfect classification in both situations throughout the entire time frame.  These results extend those found in \citet{Bachl2013Bayesian} and justify our use of the LSM emissivity modeling approach in \eqref{eq:IGARSS2013LP} on these data. 

\begin{figure}[t]
  \centering
  \subfigure[Time (pristine)]{
    \label{dustPlume}
    \includegraphics[width=0.32\linewidth]{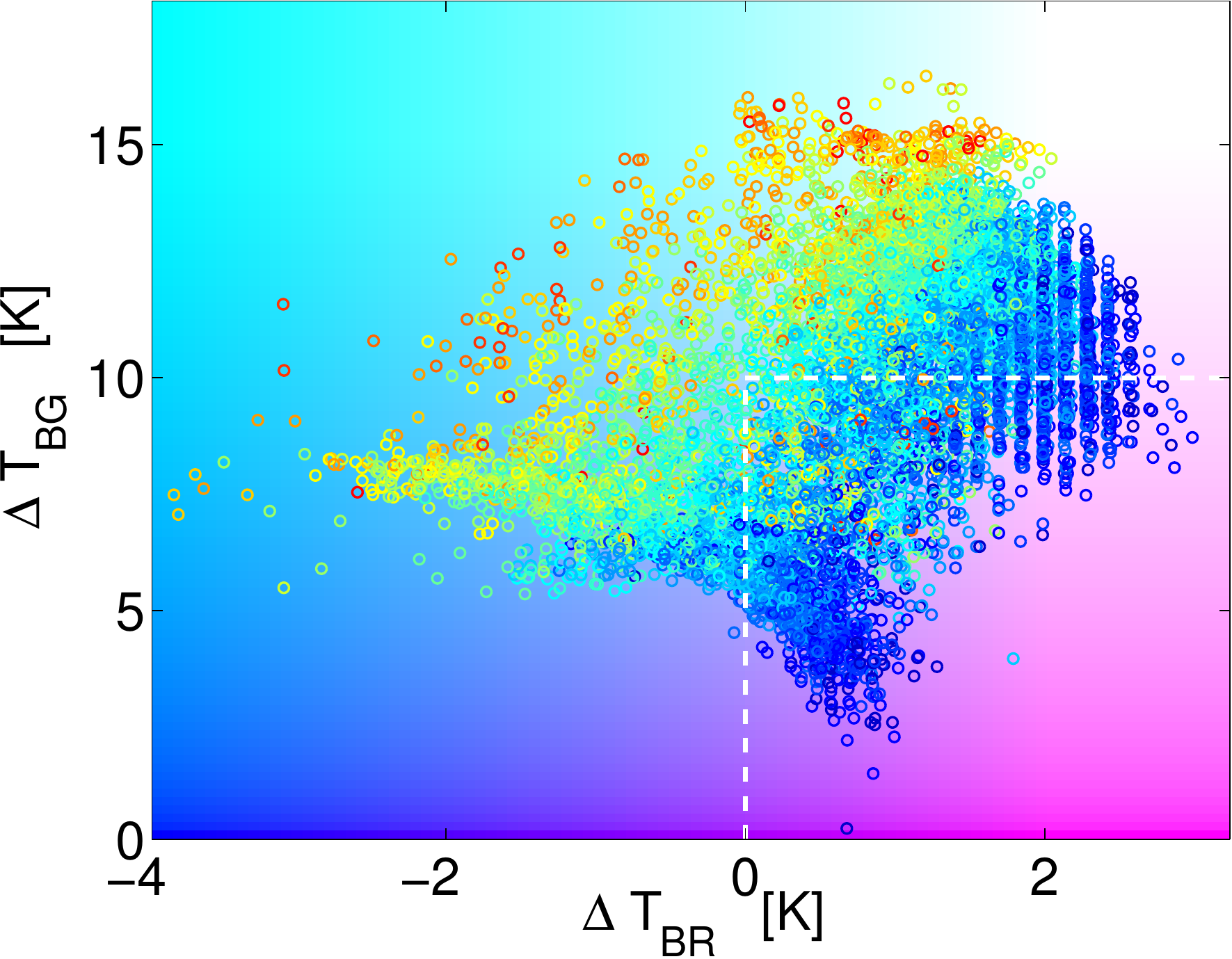} 
  }
  \subfigure[Time (dusty)]{
    \label{dustPlume}
    \includegraphics[width=0.32\linewidth]{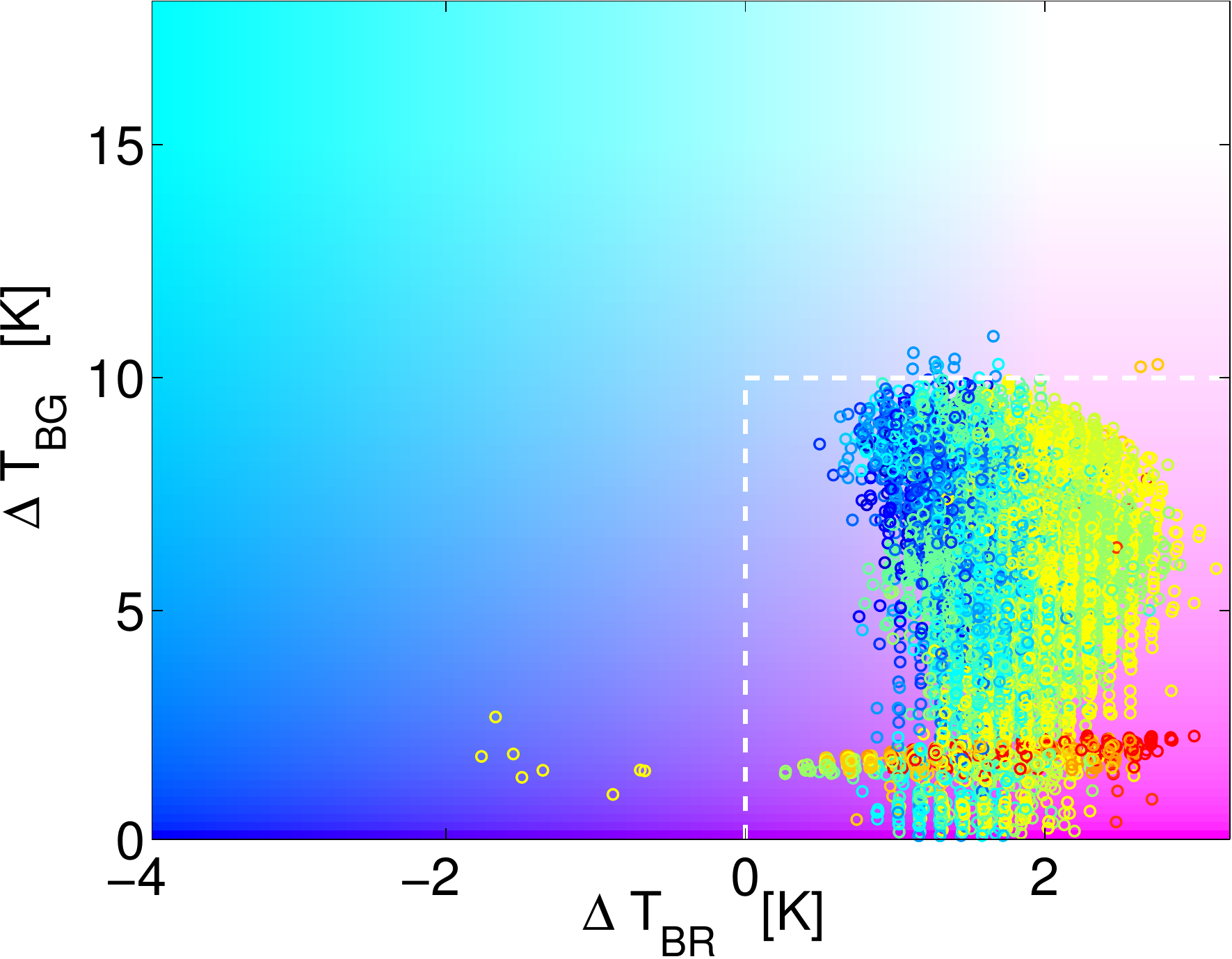} 
  }

  \subfigure[Emissivitiy (pristine)]{
    \label{sammplesEmissPristine}
    \includegraphics[width=0.32\linewidth]{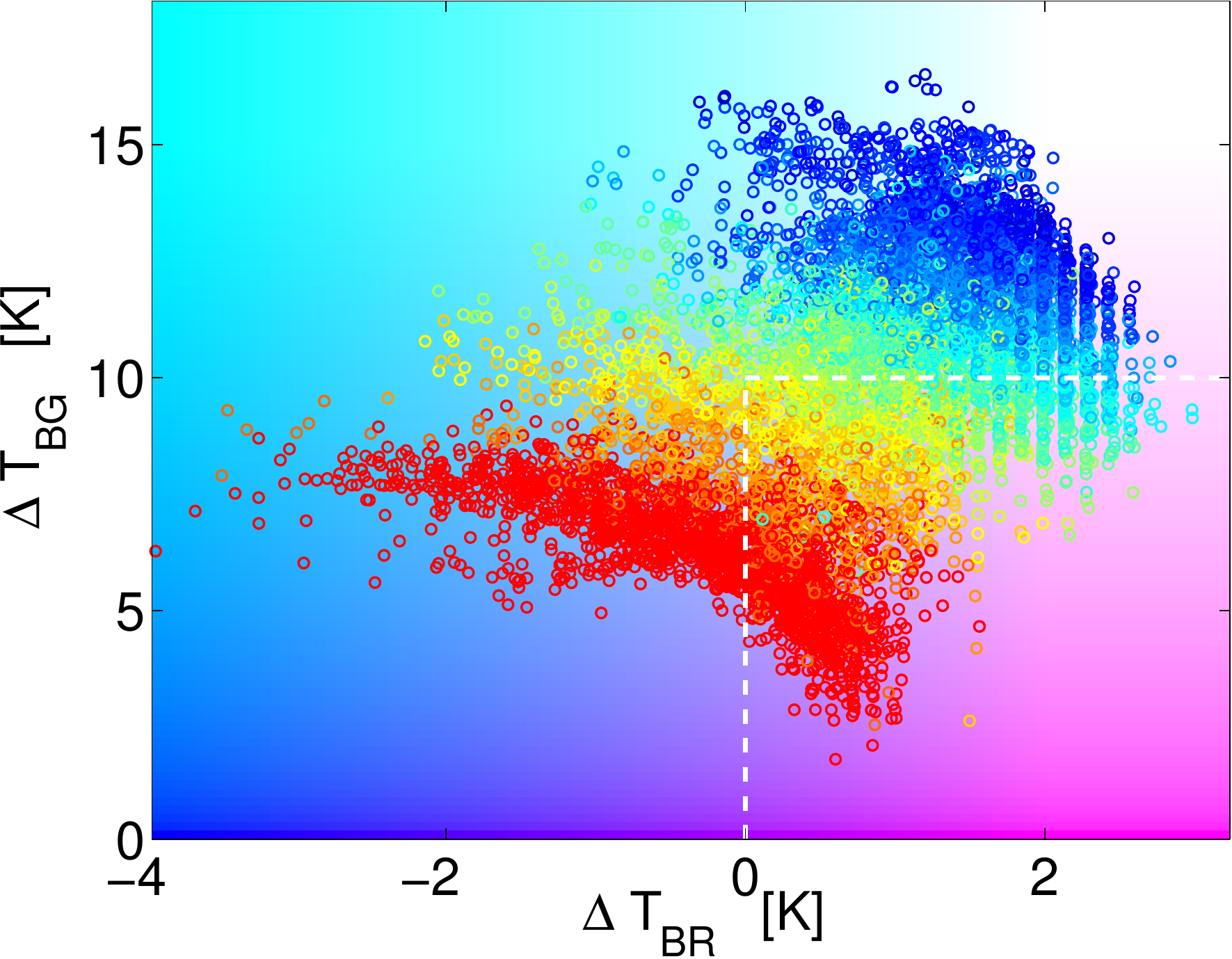} 
  }
    \subfigure[Emissivity (dusty)]{
    \label{dustPlume}
    \includegraphics[width=0.32\linewidth]{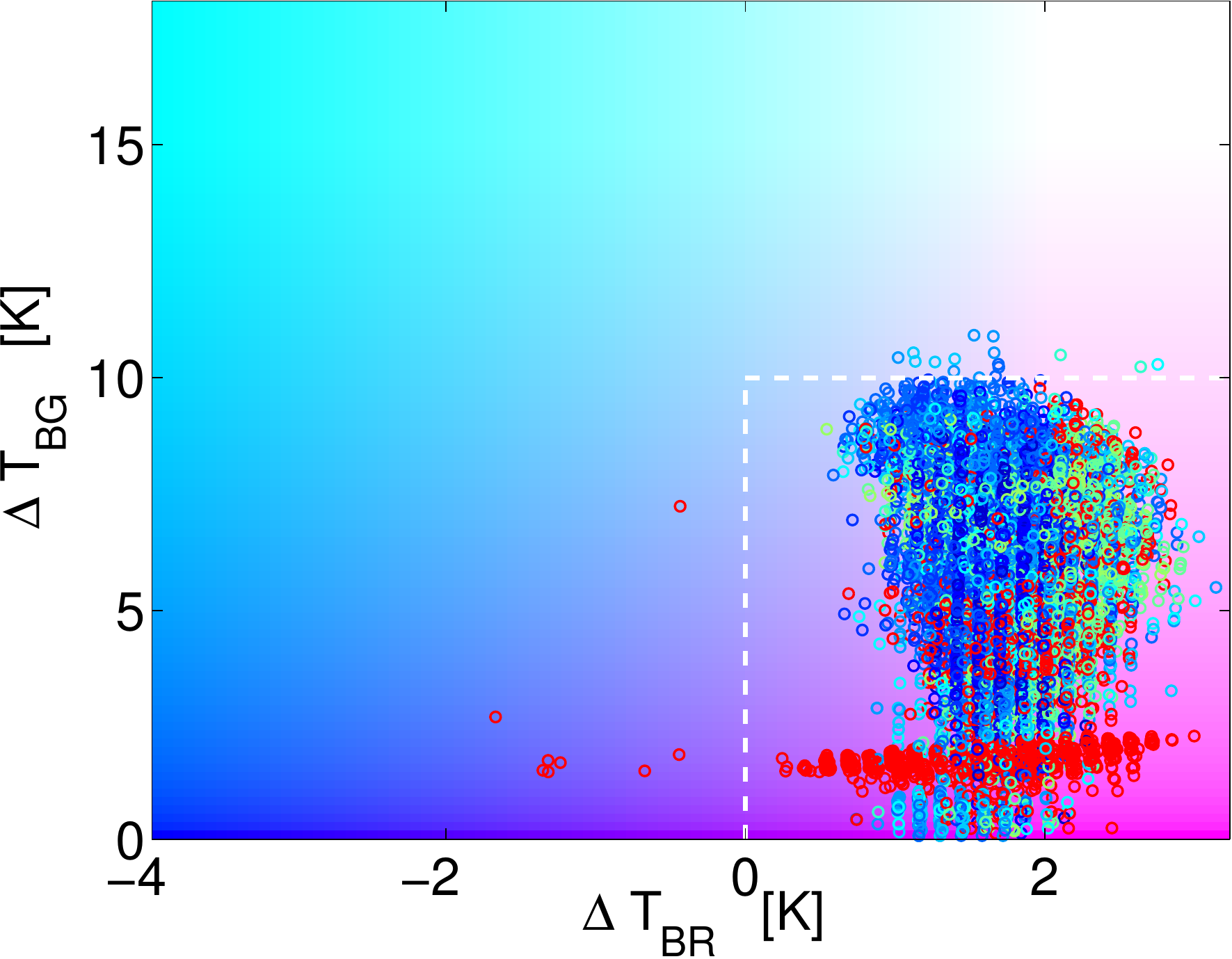} 
  }
  \caption{Green channel intensity ($x$-axis) versus red channel intensity ($y$-axis) of the labeled training data. Left column: pixels labeled as pristine; right column: pixels labeled as dusty; top row: points are colored by local time of the day, from blue (early) to red (later); bottom row: points are colored by emissivity, from blue (low) to red (high).  The white dashed lines indicate the ``no10.8'' thresholding of \cite{Ashpole2012automated}. As the entire data set is very large, each plot shows a random subsample of the full data set.}
  \label{fig:LDAcomp}
\end{figure}

Figure~\ref{fig:LDAcomp} provides some indication of why LSM improves over LDA and thresholding.  In this figure, the left column shows pixels labeled as clear, or pristine, while the right hand column pertains to dust-filled pixels.  In each figure, points are placed relative to their green channel intensity ($x$-axis) and red channel intensity ($y$-axis).  Dotted lines show the thresholding cut-offs of \citet{Ashpole2012automated}.  From the dotted lines, we immediately see why the thresholding approach performs poorly at classifying clear pixels--a large portion are inside the threshold.  

The data displayed in Figure~\ref{fig:LDAcomp} also demonstrates why LDA alone performs poorly in the early hours.  In the first row points are colored according to the local time at which the data was collected with earlier time points shown in blue.  As we can see, the red and green channel intensities for both dusty and clear points are initially very similar, while subsequently the intensities begin to diverge.  Since the LDA method classifies the data based on these intensities only, it struggles in the early hours while it improves significantly as the day progresses.  The emissivity information in the data is displayed in the bottom row of Figure~\ref{fig:LDAcomp}. For clear pixels there is a strong relationship between green and red channel intensity and emissivity levels.  By contrast, for dusty pixels, the emissivity has no relation to channel intensity since strong dust events completely block $8.3 \mu m $ radiation.  In combining this information with channel intensity in the LSM approach, we thus achieve an improved classification in the early-morning data.

\subsection{Simulation Study: Aerosol Flow}\label{sec:apps_flow}

\begin{figure*}[h]
  \centering
  \subfigure{
    \includegraphics[width=0.22\linewidth]{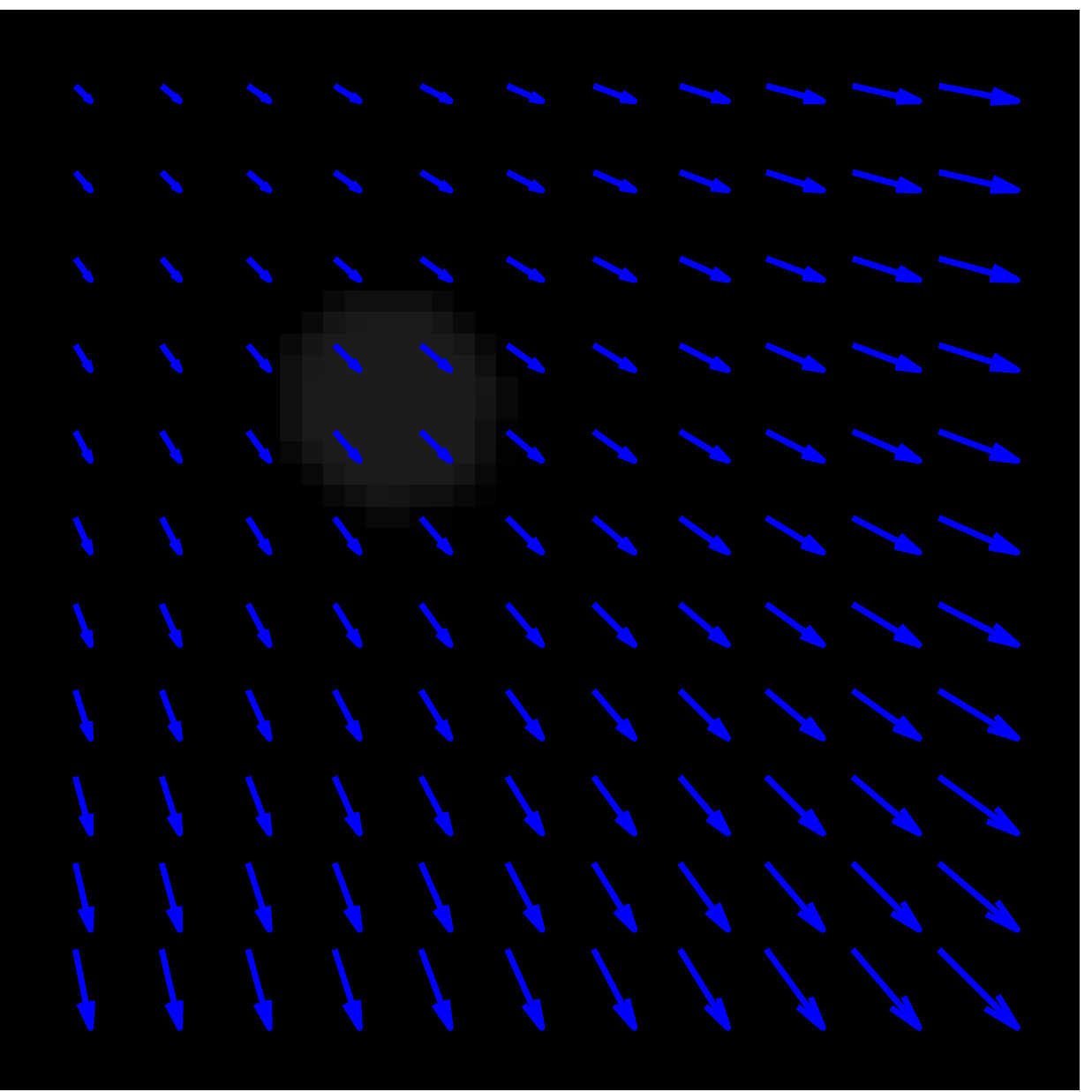}
    \includegraphics[width=0.22\linewidth]{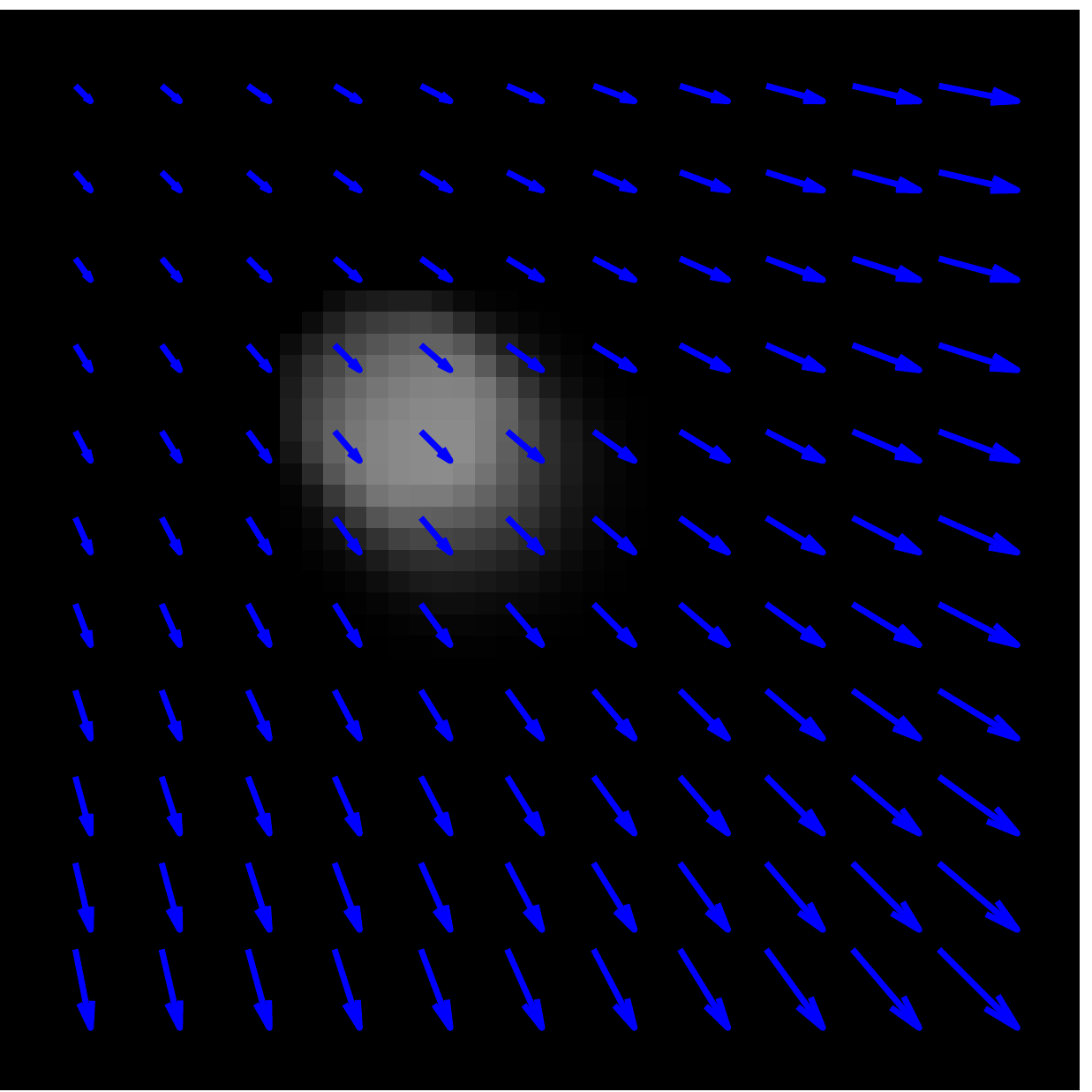}
    \includegraphics[width=0.22\linewidth]{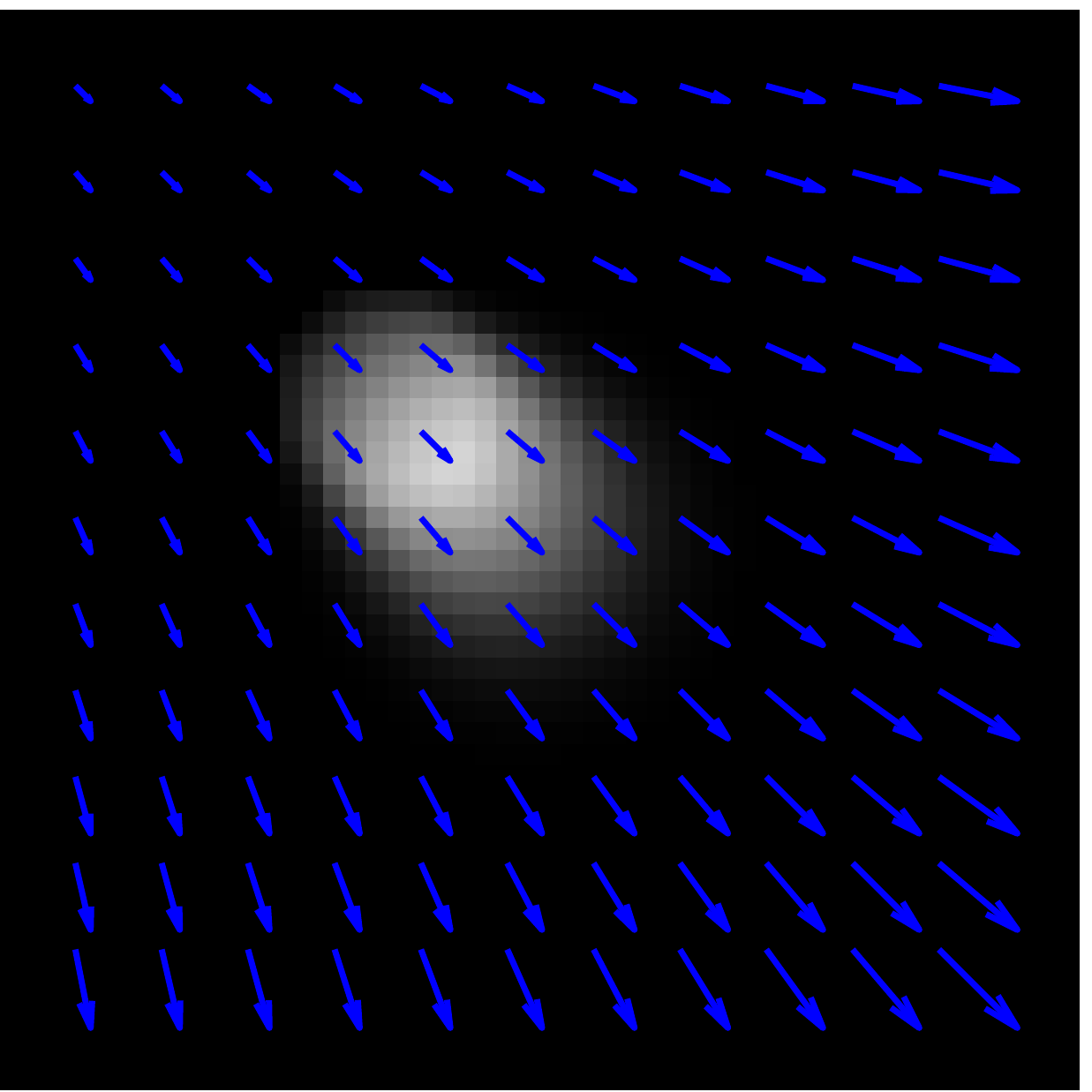}
    \includegraphics[width=0.22\linewidth]{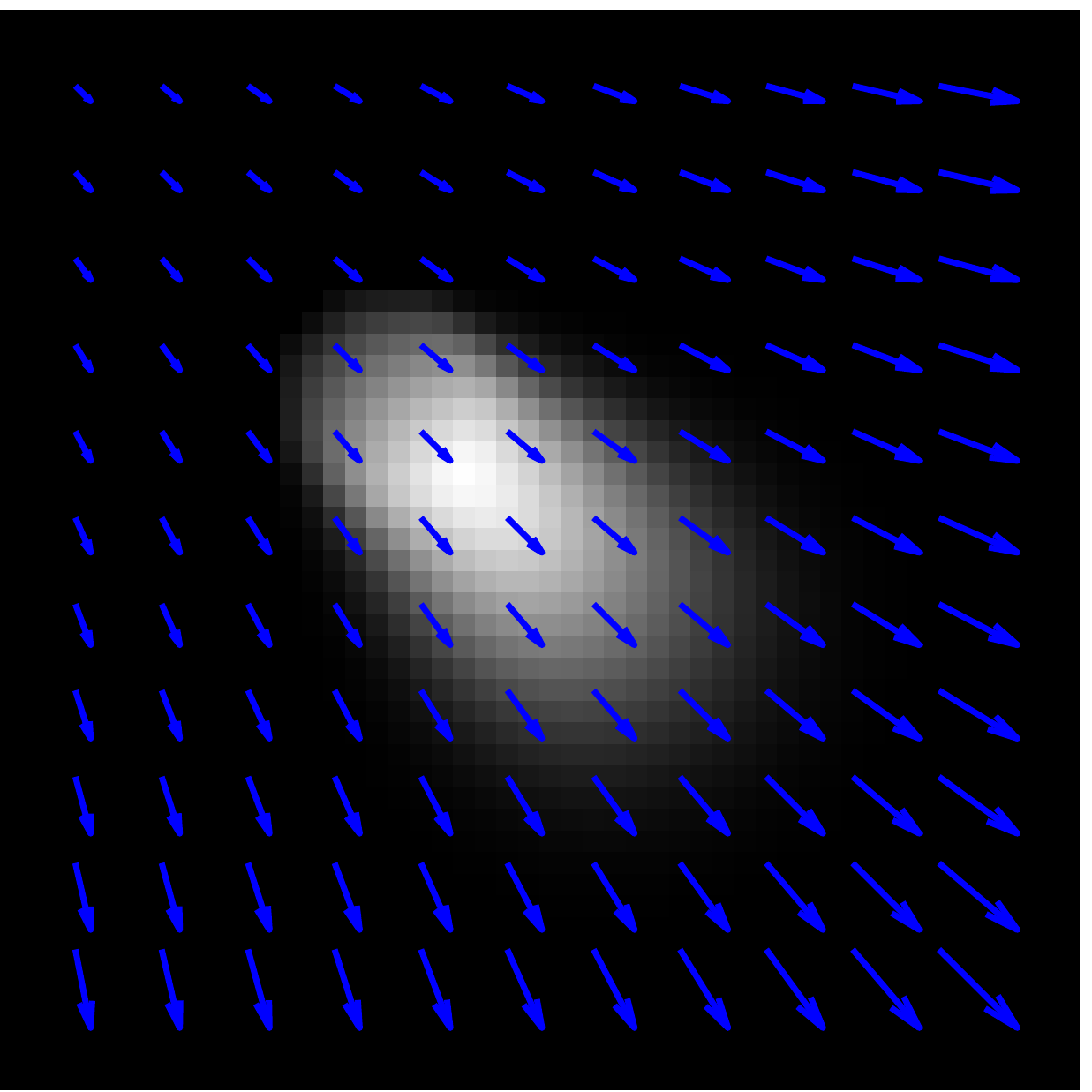}
   }
  \caption{Synthetic image sequence of a dust plume and aerosol flow.}
  \label{synthSeq}
\end{figure*}

We now compare the HS method to the ICE method in reconstructing a flow field, both under classical and the proposed Bayesian perspective.  Since ground truth is unavailable for the Saharan dust storms, we use a synthetic image sequence to illustrate the difference between the two approaches.  Figure~\ref{synthSeq} shows the progression we consider, a constant flow field with a growing dust plume.

\begin{figure*}[t]
  \centering
  \subfigure[Angular error]{
    \label{PlumeJan18}
    \includegraphics[height=4cm]{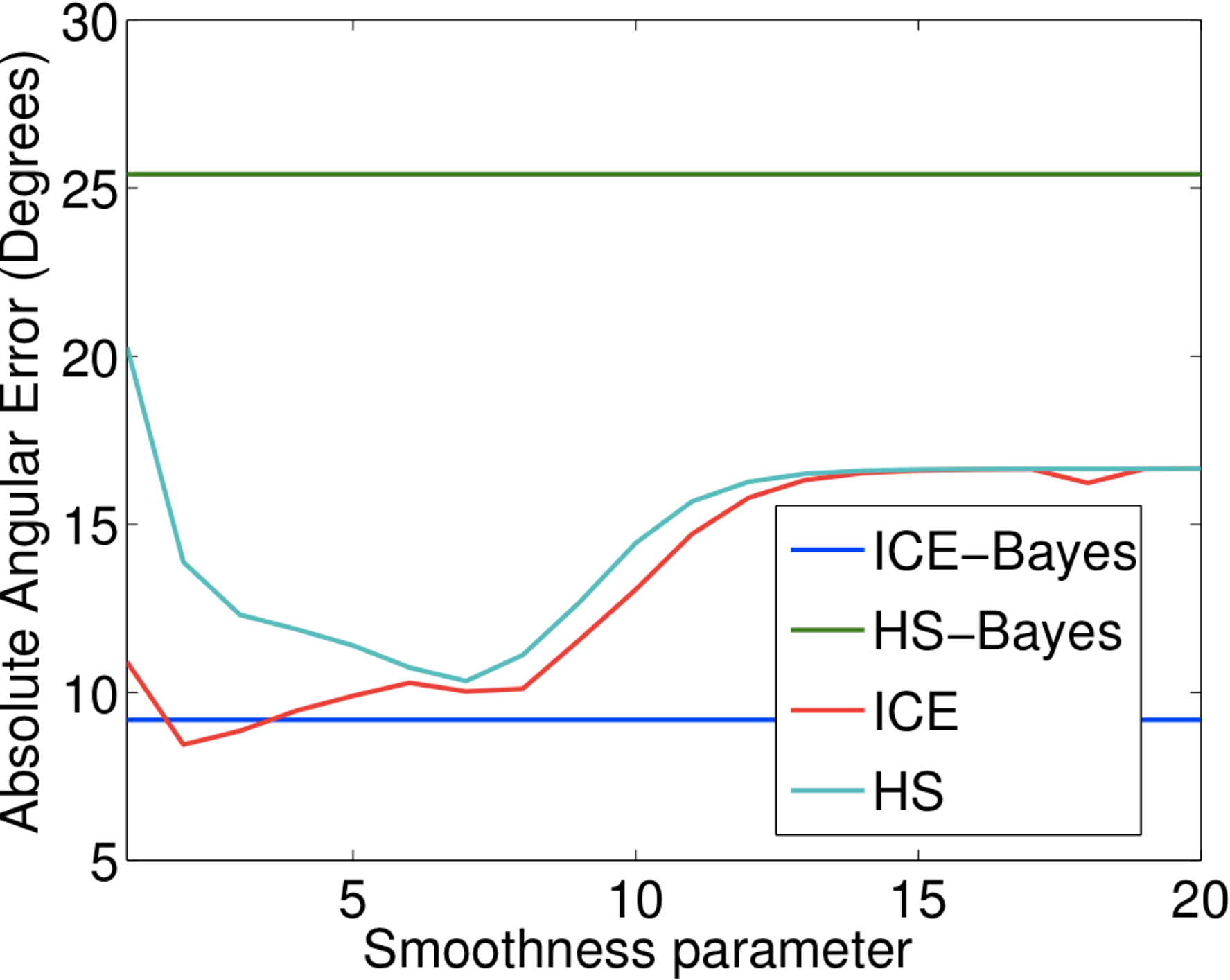} 
  }
  \subfigure[Magnitude error]{
    \label{PlumeJan18}
    \includegraphics[height=4.05cm]{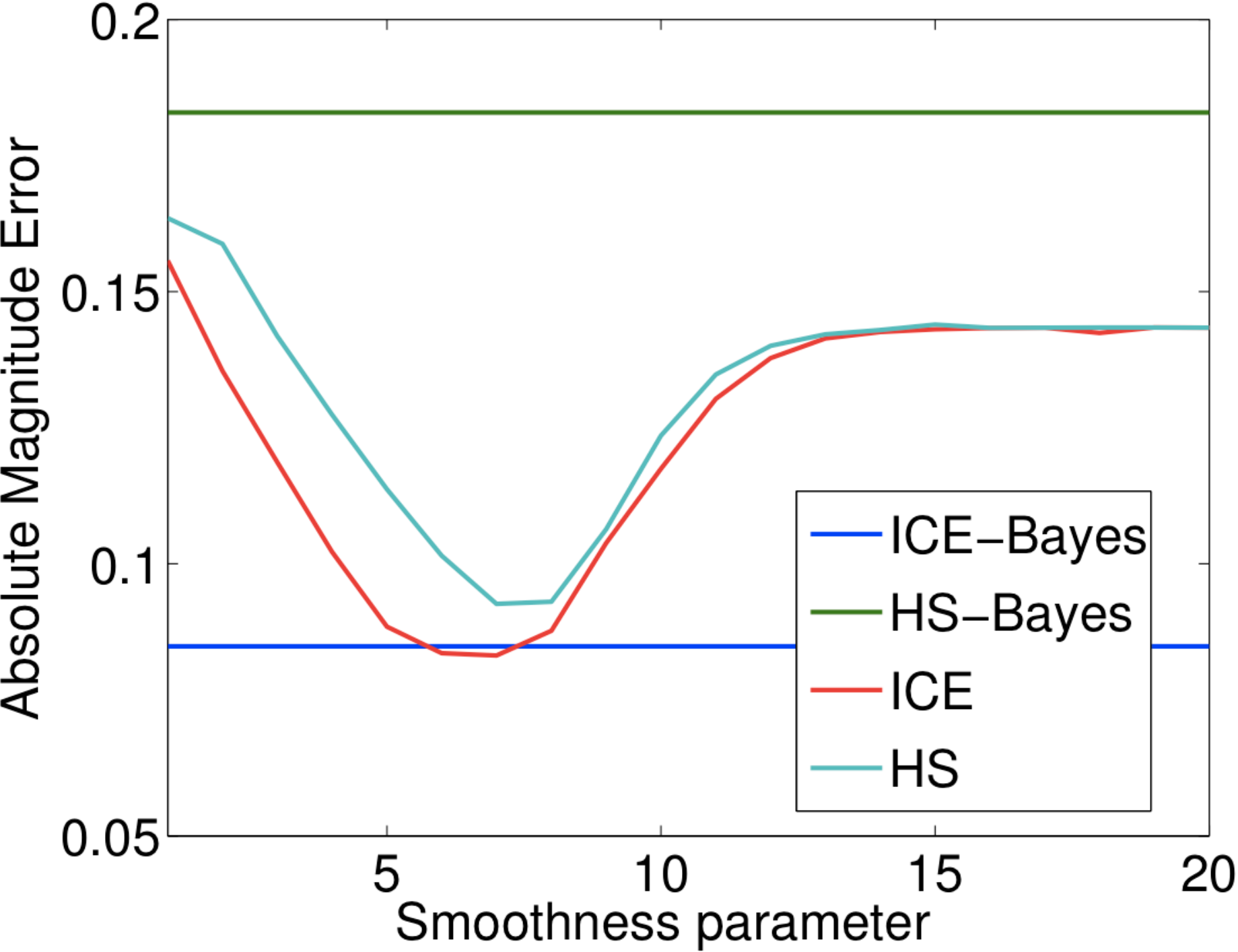} 
  }        
  \caption{Quantification of (a) absolute angular and (b) absolute magnitude error of aerosol flow estimation for the synthetic image sequence in Figure~\ref{synthSeq}.  The plots compare the errors of the ICE and the HS methods under both standard and Bayesian inference as a function of the smoothness parameter $\alpha$.}
  \label{AngAndMagErrComparison}
\end{figure*}

We assume the location of the dust plume is known and estimate the flow field under HS and ICE based on this sequence.  Figure~\ref{AngAndMagErrComparison} shows the mean absolute error in angular (left panel) and magnitude (right panel) estimates for four approaches: ICE and HS where the precision parameter $\alpha$ is set by hand (equivalent to the current best practices) and the corresponding Bayesian approaches where the INLA methodology is used to estimate this parameter.  Figure~\ref{AngAndMagErrComparison} shows several interesting features.  The first is that for any level of $\alpha$ and any error metric, the ICE approach outperforms the HS approach.  This indicates the benefit of using ICE over the BCE when the preservation of brightness assumption is clearly violated. 

The second conclusion speaks to the benefit of estimating $\alpha$ via Bayesian methods. In this context we see that, depending on the metric, different choices of $\alpha$ are optimal in case of ICE.  However, by intrinsic parameter integration, the ICE method under Bayesian estimation outperforms the regular ICE approach for almost all levels of $\alpha$, and even at its best the standard ICE method is barely better than the Bayesian approach.  Finally, there is an interesting warning regarding model misspecification.  We see that the HS method, when estimated by Bayesian methods, performs considerably worse than all other approaches, due to the violation of BCE in our example.

\subsection{Case Studies: Detection and Flow Estimation}\label{sec:apps_case}

\begin{figure*}[p]
  \centering
  \subfigure{
    \label{PlumeJan_08_1}
    \includegraphics[width=\cswidth,height=\csheight]{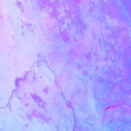} 
  } 
  \subfigure{
    \label{PlumeJan_08_2}
    \includegraphics[width=\cswidth,height=\csheight]{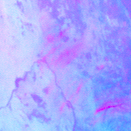} 
  }
  \subfigure{
    \label{PlumeJan_08_3}
    \includegraphics[width=\cswidth,height=\csheight]{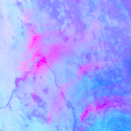} 
  }
  \subfigure{
    \label{PlumeJan_08_1_flowHS}
    \includegraphics[width=\cswidth,height=\csheight]{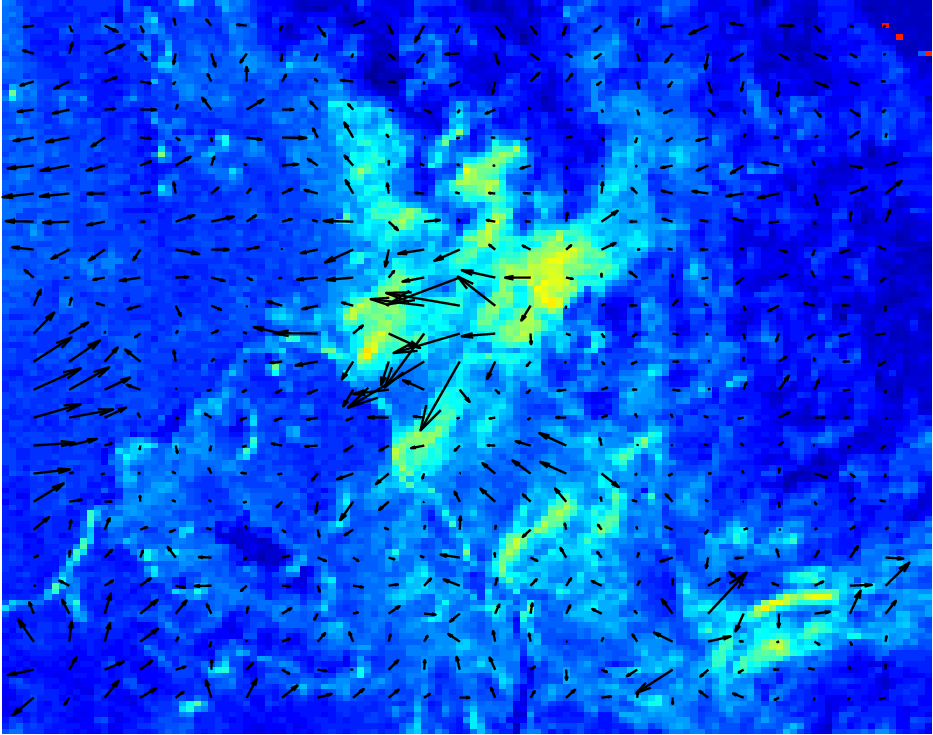} 
  } 
  \subfigure{
    \label{PlumeJan_08_2_flowHS}
    \includegraphics[width=\cswidth,height=\csheight]{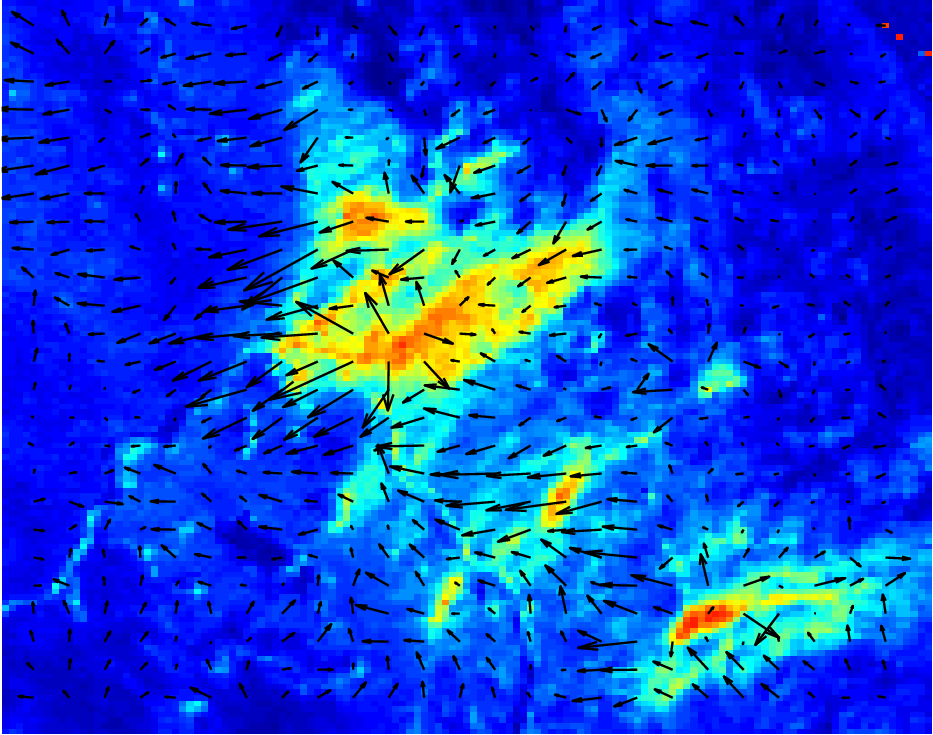} 
  }
  \subfigure{
    \label{PlumeJan_08_3_flowHS}
    \includegraphics[width=\cswidth,height=\csheight]{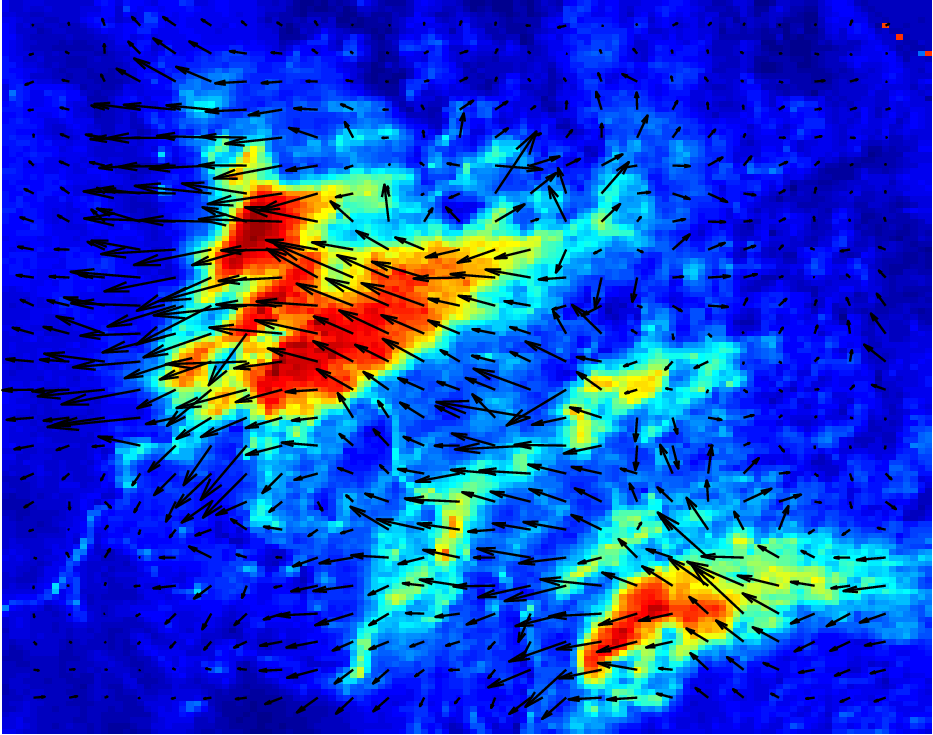} 
  }
  
  \subfigure{
    \label{PlumeJan_08_1_flowICE}
    \includegraphics[width=\cswidth,height=\csheight]{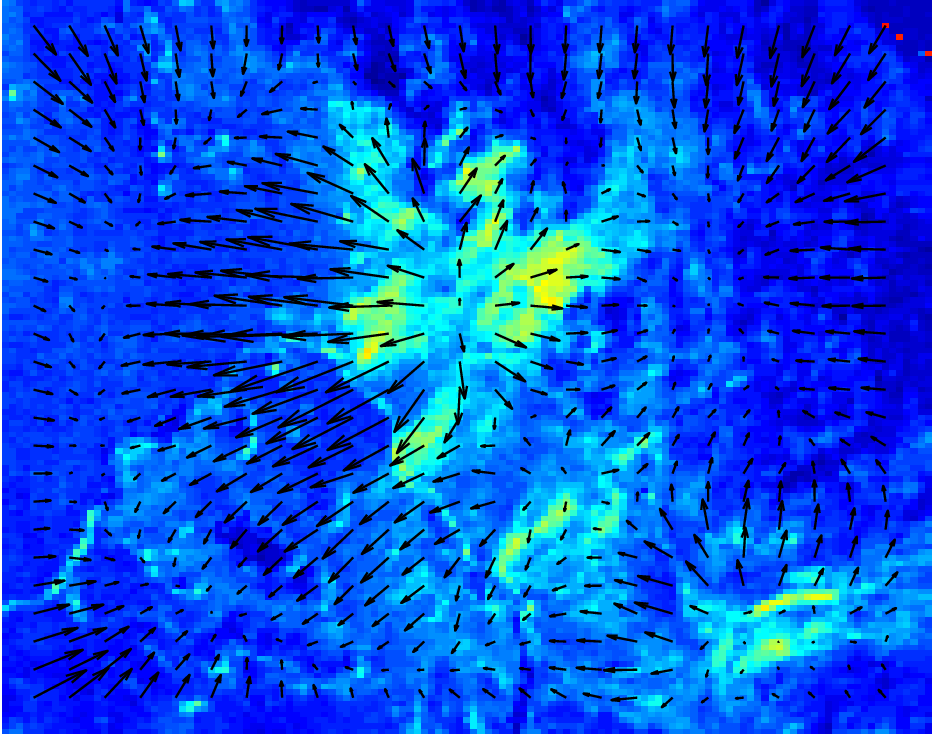} 
  } 
  \subfigure{
    \label{PlumeJan_08_2_flowICE}
    \includegraphics[width=\cswidth,height=\csheight]{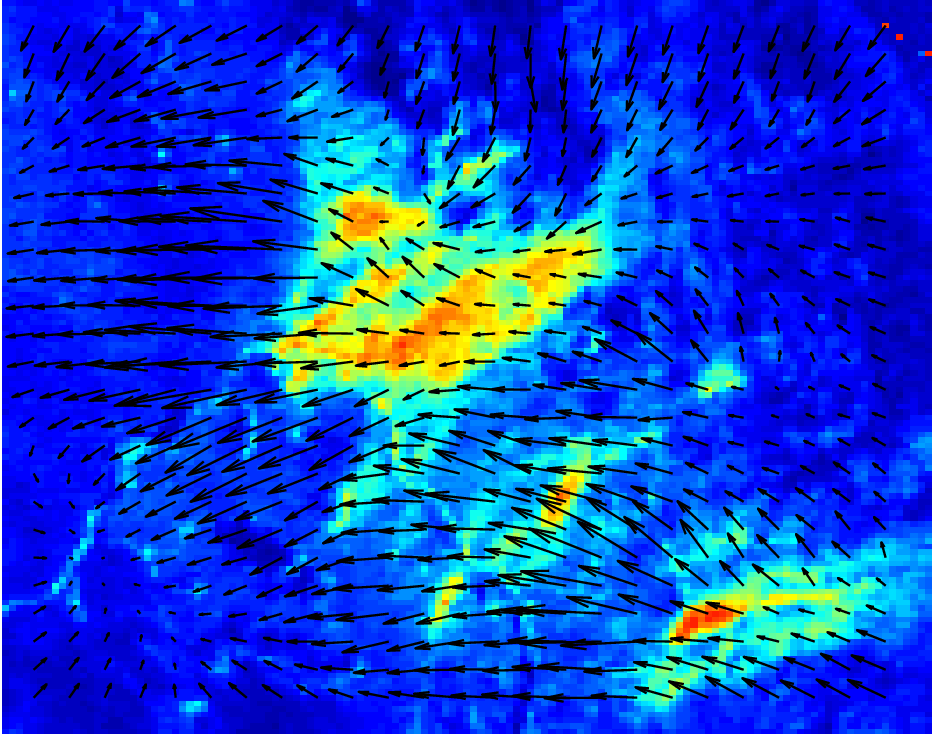} 
  }
  \subfigure{
    \label{PlumeJan_08_3_flowICE}
    \includegraphics[width=\cswidth,height=\csheight]{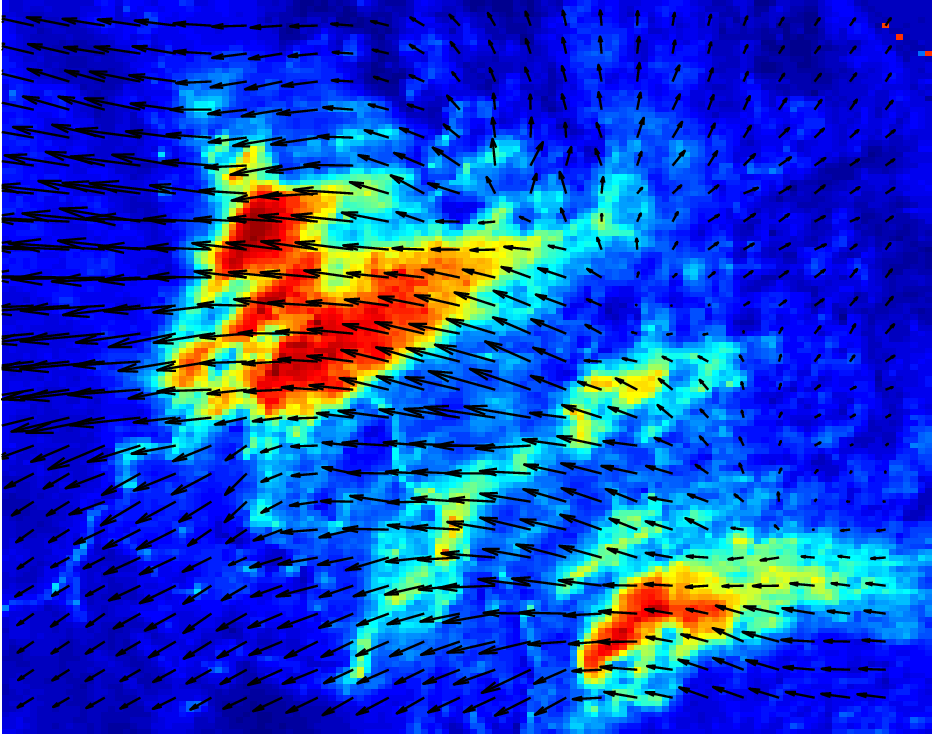} 
  }
  \caption{Dust plume on January 8, 2010 at 7.15 am, 8.30 am and 11 am GMT. Top row: observed satellite data in false color; middle row: pixel-wise LSM probability of dust estimates (with high probabilities in red) overlaid with the Bayesian HS flow field; bottom row: same pixel-wise probability of dust estimates as above now overlaid with the Bayesian ICE flow field.}
  \label{fig:caseStudy}
\end{figure*}
\begin{figure*}[p]
 \centering
 \subfigure{
    \label{PlumeJan_16_1}
    \includegraphics[width=\cswidth,height=\csheight]{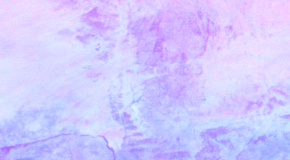} 
  } 
  \subfigure{
    \label{PlumeJan_16_2}
    \includegraphics[width=\cswidth,height=\csheight]{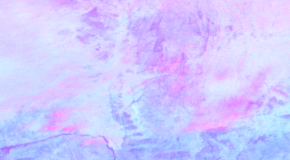} 
  }
  \subfigure{
    \label{PlumeJan_16_3}
    \includegraphics[width=\cswidth,height=\csheight]{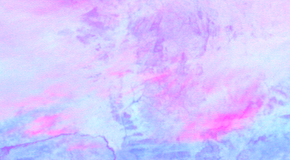} 
  }
  
  \subfigure{
    \label{PlumeJan_16_3_flowHS}
    \includegraphics[width=\cswidth,height=\csheight]{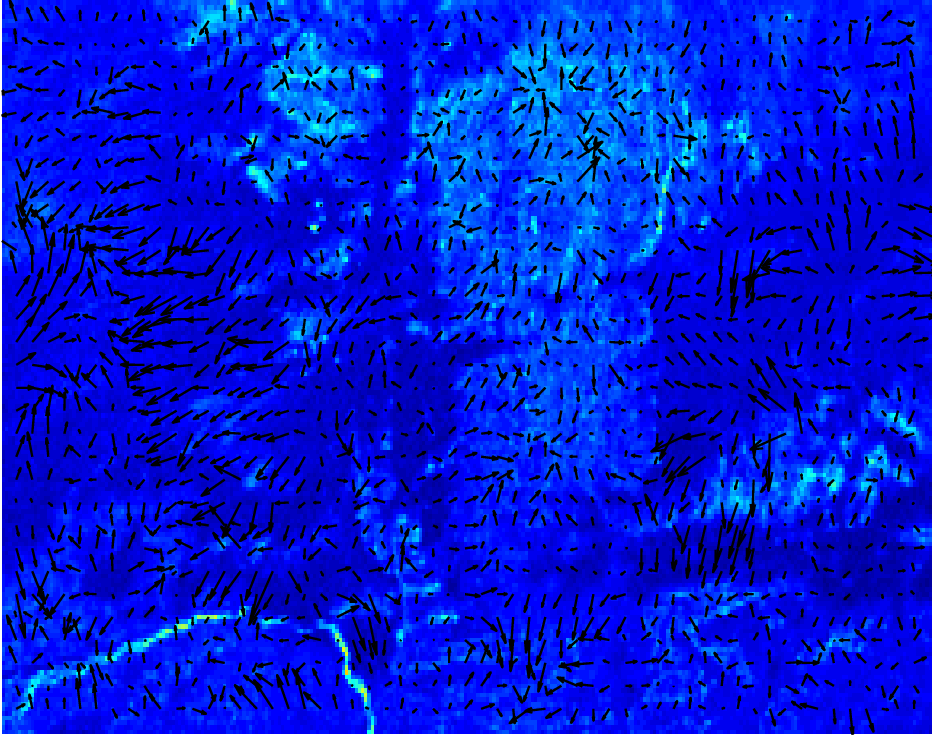} 
  }  
  \subfigure{
    \label{PlumeJan_16_3_flowHS}
    \includegraphics[width=\cswidth,height=\csheight]{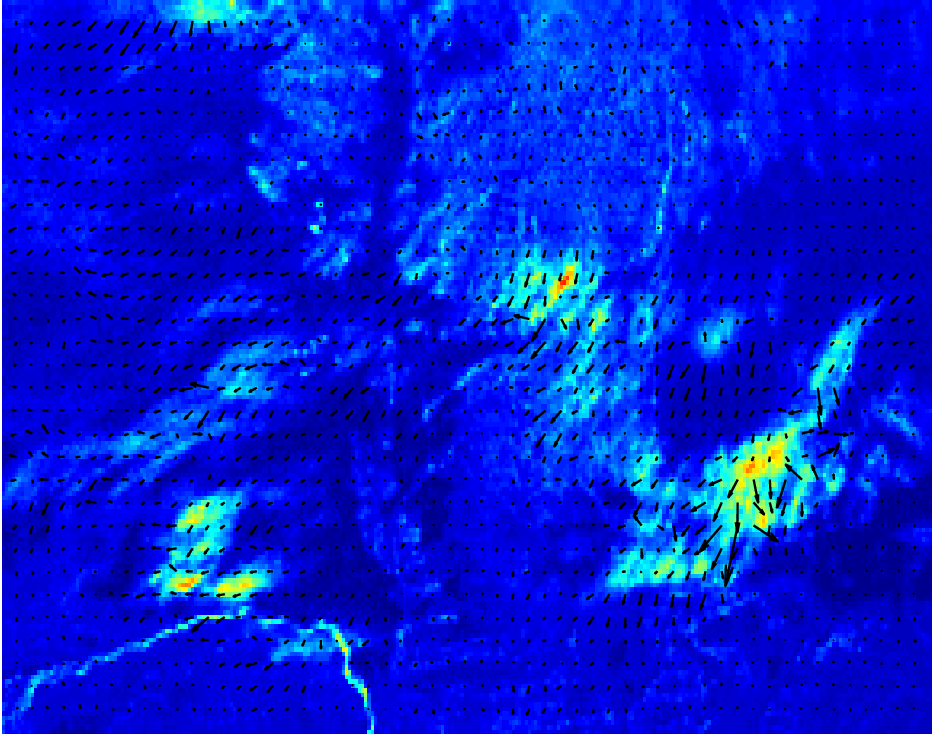} 
  } 
  \subfigure{
    \label{PlumeJan_16_3_flowHS}
    \includegraphics[width=\cswidth,height=\csheight]{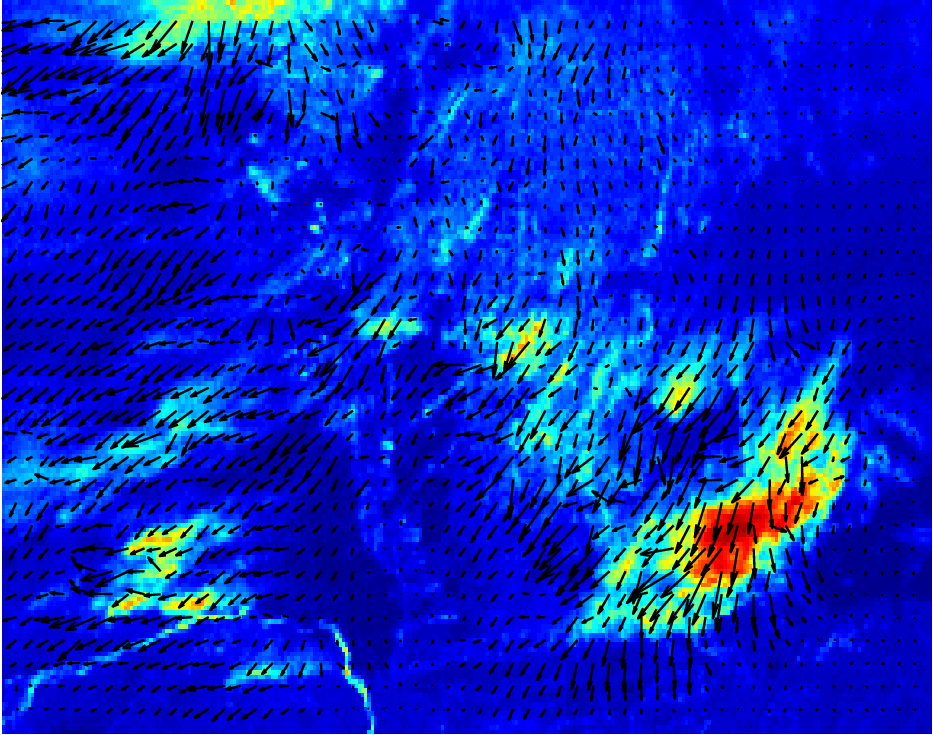} 
  }
    \subfigure{
    \label{PlumeJan_16_3_flowICE}
    \includegraphics[width=\cswidth,height=\csheight]{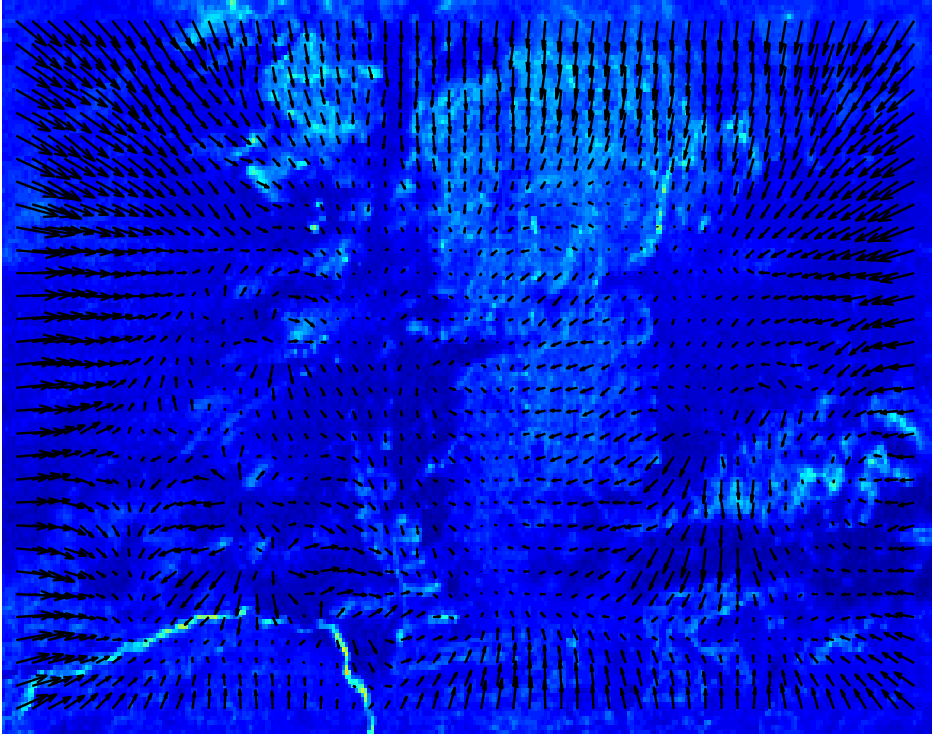} 
  }  
  \subfigure{
    \label{PlumeJan_16_3_flowICE}
    \includegraphics[width=\cswidth,height=\csheight]{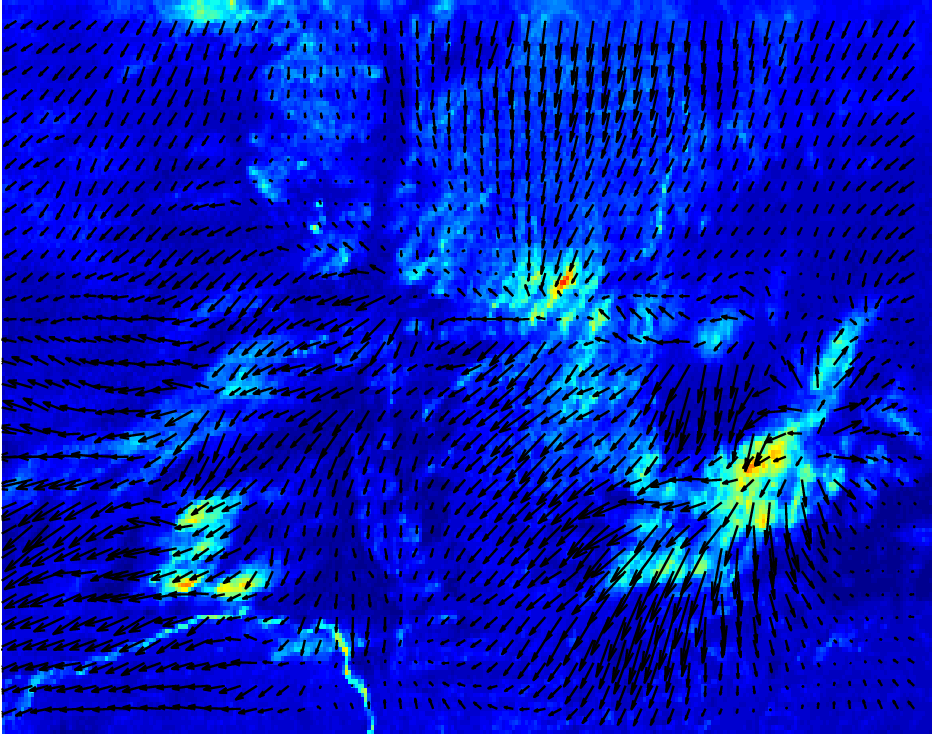} 
  } 
  \subfigure{
    \label{PlumeJan_16_3_flowICE}
    \includegraphics[width=\cswidth,height=\csheight]{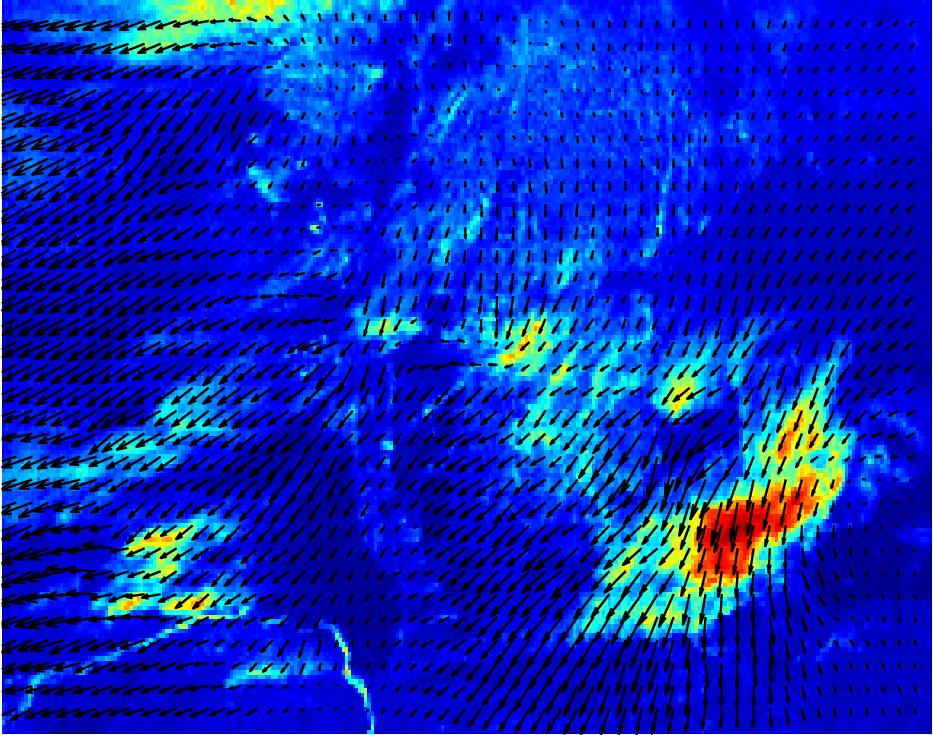} 
  }
     
  \caption{Dust plume on January 16, 2010 at 10.15 am, 11.45 am and 1 pm GMT. Top row: observed satellite data in false color; middle row: pixel-wise LSM probability of dust estimates overlaid with the Bayesian HS flow field; bottom row: same pixel-wise probability of dust estimates as above now overlaid with the Bayesian ICE flow field.}
  \label{fig:caseStudy2}
\end{figure*}

After establishing the good performance of our dust detection routine and the Bayesian ICE method of reconstructing the flow field, we highlight the use of our framework during the evolution of two separate dust storms.  Figures~\ref{fig:caseStudy} and~\ref{fig:caseStudy2} show dust storms that occurred during January 8, 2010 and January 16, 2010, respectively.  The figures show the pixel-wise probability of dust estimates under the emissivity LSM approach and, furthermore, compare the estimated flow fields under the Bayesian HS and the Bayesian ICE approaches. 

We see several features from Figures~\ref{fig:caseStudy} and~\ref{fig:caseStudy2}.  The first is that the detection appears to be working well.  Points which are clearly dusty are correctly given high probabilities, while the model captures uncertainty in the estimates around the edges of the dust plumes.  Secondly, we see why the ICE method is preferred over standard HS.  There is considerably more regularity to the estimated flow field in the third row than the second row, especially in the first two time points.  This enables a coherent reconstruction of the dust plume flow.  Furthermore, the Bayesian HS method seems unable to detect the flow of smaller dust storms, such as the one featured in the lower right hand corner of the plots in Figure~\ref{fig:caseStudy}.

\section{Discussion}
We have outlined a Bayesian framework leveraging the recently developed INLA methodology for detecting and tracking dust storms.  The approach makes several developments, including a superior dust detection methodology, a link between the classical literature of optical flow and GMRFs--which incidentally shows how Bayesian estimation can alleviate issues related to the setting of tuning parameters--and the use of the ICE to model flow fields where an assumption of brightness constancy is inconsistent with the physical process.  Simulation studies have shown the improved performance of both our storm detection framework and the Bayesian ICE model over existing procedures and real world examples have shown the implications of this improvement.

Considerable work remains, both from the application and methodological perspectives.  The model for $\eta$ appears to work quite well in our current data, but it could be extended in several obvious manners.  The most useful of these would be  to make the estimates of $\eta$ depend not just on emissivity and image intensity, but to also include spatial and temporal dependence on neighboring estimates.  In practice this appeared to be unnecessary in our current approach--and the computational challenges to such estimation proved challenging--however as the performance of the INLA software continually improves, such developments may become helpful. Another worthwhile extension would be to take the local time or other covariates such as satellite viewing angle of a particular pixel location into account. In particular if the dust analysis is extended from the forenoon to a whole day the former might be a critical feature to prevent a degradation of detection performance.

Regarding our procedure for estimating flow fields, a general smoothness assumption or even local constancy as within the \cite{Lucas1981iterative} approach can in most cases be justified, it is particularly appealing that work like that of \cite{Lindgren2011explicit} as well as \cite{Simpson2012In} reveals links between Gaussian fields and GMRFs via stochastic partial differential equations (SPDEs). Thus, future work may find this link as a mode to refine the prior of the GMRF in terms of expressing a transport phenomenon via its SPDE and thereby gain further insight into how it is reflected by the given data. 

The connection to continuously modeled phenomena also comes up at the methodological intersection with image processing methods. Traditionally, inferring the HS optical flow was subject to solving a variational formulation of the problem via the corresponding Euler-Lagrange equations. Most importantly, the variational perspective leads to further insight about the properness of the resulting GMRF with respect to the function space the data are sampled from. As shown by \cite{Schnorr1991Determining}, relatively mild conditions, namely a mildly restricted Sobolev space, are sufficient to guarantee this properness. It should also be mentioned that the likelihood term and respective choices of the error penalty of the HS optical flow and related methods has consistently been subject to several studies. Here, the corresponding flexibility of the GLM formulation and the INLA methodology might excel in further in-depth analyses.

While showing that remote sensing equipment can be used to detect and track dust storms was our initial goal, there are considerable applied advances that can now be pursued.  This related to projecting the dust storm into the future, as well as ``rewinding'' the storm to pinpoint its source.  The advantage of our statistical approach is that it inherently enables the uncertainty of such assessments to be expressed.  This, in turn, will allow us to issue probabilistic forecasts and leverage the recent work in forecasting methodology \citep{Gneiting2005Weather, Schefzik&2013}.  Such probabilistic forecasts would be of considerable interest to the Earth observation community and could also be fed into larger models of global transport phenomena.
\section*{Acknowledgments}

This work has been supported by the German Research Foundation (DFG) within the programme "Spatio/Temporal Graphical Models and Applications in Image Analysis", grant GRK 1653.  Alex Lenkoski and Thordis Thorarinsdottir's work is also supported by Statistics for Innovation $(sfi)^2$ in Oslo.




\bibliography{FabianGraphMod,morebibs}

\begin{thebibliography}{}

\bibitem[Aberg et~al., 2005]{Aberg2005image}
Aberg, S., Lindgren, F., Malmberg, A., Holst, J., and Holst, U. (2005).
\newblock An image warping approach to spatio-temporal modelling.
\newblock {\em Environmetrics}, 16(8):833--848.

\bibitem[Ashpole and Washington, 2012]{Ashpole2012automated}
Ashpole, I. and Washington, R. (2012).
\newblock An automated dust detection using seviri: A multiyear climatology of
  summertime dustiness in the central and western sahara.
\newblock {\em Journal of Geophysical Research: Atmospheres}, 117(D8):D08202.

\bibitem[Bachl et~al., 2012]{Bachl2012Bayesian}
Bachl, F.~E., Fieguth, P., and Garbe, C.~S. (2012).
\newblock A bayesian approach to spaceborn hyperspectral optical flow
  estimation on dust aerosols.
\newblock In {\em Proceedings of the International Geoscience and Remote
  Sensing Symposium 2012}, pages 256--259.

\bibitem[Bachl et~al., 2013]{Bachl2013Bayesian}
Bachl, F.~E., Fieguth, P., and Garbe, C.~S. (2013).
\newblock Bayesian inference on integrated continuity fluid flows and their
  application to dust aerosols.
\newblock In {\em Proceedings of the International Geoscience and Remote
  Sensing Symposium 2013}, page to appear.

\bibitem[Bachl and Garbe, 2012]{Bachl2012Classify}
Bachl, F.~E. and Garbe, C.~S. (2012).
\newblock Classifying and tracking dust plumes from passive remote sensing.
\newblock In {Ouwehand}, L., editor, {\em Proceedings of the ESA, SOLAS \& EGU
  Joint Conference 'Earth Observation for Ocean-Atmosphere Interaction
  Science'}, volume 703 of {\em ESA Special Publication}, pages S1--3,
  Frascati, Italy. European Space Agency, European Space Agency Communications.

\bibitem[Besag, 1974]{Besag1974Spatial}
Besag, J. (1974).
\newblock Spatial interaction and the statistical analysis of lattice systems
  (with discussion).
\newblock {\em Journal of the Royal Statistical Society Series B}, 36:192--236.

\bibitem[Brindley et~al., 2012]{Brindley2012critical}
Brindley, H., Knippertz, P., Ryder, C., and Ashpole, I. (2012).
\newblock A critical evaluation of the ability of the spinning enhanced visible
  and infrared imager (seviri) thermal infrared red-green-blue rendering to
  identify dust events: Theoretical analysis.
\newblock {\em Journal of Geophysical Research: Atmospheres}, 117(D7):D07201.

\bibitem[Corpetti et~al., 2002]{Corpetti2002Dense}
Corpetti, T., Memin, E., and Perez, P. (2002).
\newblock Dense estimation of fluid flows.
\newblock {\em IEEE Transactions on Pattern Analysis and Machine Intelligence},
  24(3):365--380.

\bibitem[Eissa et~al., 2012]{Eissa2012Dust}
Eissa, Y., Ghedira, H., Ouarda, T. B. M.~J., and Chiesa, M. (2012).
\newblock Dust detection over bright surfaces using high-resolution visible
  seviri images.
\newblock In {\em Proceedings of the International Geoscience and Remote
  Sensing Symposium 2012}, pages 3674--3677.

\bibitem[Gilleland et~al., 2010]{Gilleland2010Analyzing}
Gilleland, E., Lindstr\"{o}m, J., and Lindgren, F. (2010).
\newblock Analyzing the image warp forecast verification method on
  precipitation fields from the icp.
\newblock {\em Weather and Forecasting}, 25(4):1249--1262.

\bibitem[Glasbey and Mardia, 1998]{Glasbey1998review}
Glasbey, C.~A. and Mardia, K.~V. (1998).
\newblock A review of image-warping methods.
\newblock {\em Journal of Applied Statistics}, 25(2):155--171.

\bibitem[Gneiting and Raftery, 2005]{Gneiting2005Weather}
Gneiting, T. and Raftery, A.~E. (2005).
\newblock Weather forecasting with ensemble methods.
\newblock {\em Science}, 310:248--249.

\bibitem[Heitz et~al., 2010]{Heitz2010Variational}
Heitz, D., M\'{e}min, E., and Schn\"{o}rr, C. (2010).
\newblock Variational fluid flow measurements from image sequences: synopsis
  and perspectives.
\newblock {\em Experiments in Fluids}, 48(3):369--393.

\bibitem[Horn and Schunck, 1981]{Horn1981Determining}
Horn, B. K.~P. and Schunck, B.~G. (1981).
\newblock Determining optical flow.
\newblock {\em Artificial Intelligence}, 17(1-3):185--203.

\bibitem[Kl\"user and Schepanski, 2009]{Klueser2009Remote}
Kl\"user, L. and Schepanski, K. (2009).
\newblock Remote sensing of mineral dust over land with msg infrared channels:
  A new bitemporal mineral dust index.
\newblock {\em Remote Sensing of Environment}, 113(9):1853--1867.

\bibitem[Krajsek and Mester, 2006a]{Krajsek2006maximum}
Krajsek, K. and Mester, R. (2006a).
\newblock A maximum likelihood estimator for choosing the regularization
  parameters in global optical flow methods.
\newblock In {\em IEEE International Conference on Image Processing}, pages
  1081--1084.

\bibitem[Krajsek and Mester, 2006b]{Krajsek2006equivalence}
Krajsek, K. and Mester, R. (2006b).
\newblock On the equivalence of variational and statistical differential motion
  estimation.
\newblock In {\em IEEE Southwest Symposium on Image Analysis and
  Interpretation, Denver, Colorado, pp. 11-15..}

\bibitem[Lensky and Rosenfeld, 2008]{Lensky2008Clouds}
Lensky, I. and Rosenfeld, D. (2008).
\newblock Clouds-aerosols-precipitation satellite analysis tool (capsat).
\newblock {\em Atmospheric Chemistry and Physics}, 8(8):6739--6753.

\bibitem[Lindgren et~al., 2011]{Lindgren2011explicit}
Lindgren, F., Rue, H., and Lindstr\"{o}m, J. (2011).
\newblock An explicit link between gaussian fields and gaussian markov random
  fields: the stochastic partial differential equation approach.
\newblock {\em Journal of the Royal Statistical Society Series B},
  73(4):423--498.

\bibitem[Lucas and Kanade, 1981]{Lucas1981iterative}
Lucas, B.~D. and Kanade, T. (1981).
\newblock An iterative image registration technique with an application to
  stereo vision.
\newblock In {\em Proceedings of the 7th international joint conference on
  Artificial intelligence}, volume~2, pages 674--679, San Francisco, CA, USA.
  Morgan Kaufmann Publishers Inc.

\bibitem[{Marzban} and {Sandgathe}, 2010]{Marzban2010Optical}
{Marzban}, C. and {Sandgathe}, S. (2010).
\newblock {Optical Flow for Verification}.
\newblock {\em Weather and Forecasting}, 25:1479--1494.

\bibitem[Rivas-Perea et~al., 2010]{Rivas-Perea2010Traditional}
Rivas-Perea, P., Rosiles, J., and Chacon, M. (2010).
\newblock Traditional and neural probabilistic multispectral image processing
  for the dust aerosol detection problem.
\newblock In {\em Image Analysis Interpretation (SSIAI), 2010 IEEE Southwest
  Symposium on}, pages 169 --172.

\bibitem[Rue and Held, 2005]{Rue2005Gaussian}
Rue, H. and Held, L. (2005).
\newblock {\em Gaussian Markov Random Fields: Theory and Applications}, volume
  104 of {\em Monographs on Statistics and Applied Probability}.
\newblock Chapman \& Hall, London.

\bibitem[Rue et~al., 2009]{Rue2009Approximate}
Rue, H., Martino, S., and Chopin, N. (2009).
\newblock Approximate bayesian inference for latent gaussian models by using
  integrated nested laplace approximations.
\newblock {\em Journal Of The Royal Statistical Society Series B},
  71(2):319--392.

\bibitem[Schefzik et~al., 2013]{Schefzik&2013}
Schefzik, R., Thorarinsdottir, T.~L., and Gneiting, T. (2013).
\newblock Uncertainty quantification in complex simulation models using
  ensemble copula coupling.
\newblock {\em Statistical Science}, in press.

\bibitem[Schepanski et~al., 2007]{Schepanski2007new}
Schepanski, K., Tegen, I., Laurent, B., Heinold, B., and Macke, A. (2007).
\newblock A new saharan dust source activation frequency map derived from
  msg-seviri ir-channels.
\newblock {\em Geophysical Research Letters}, 34(18):L13401.

\bibitem[{Schmetz} et~al., 2002]{Schmetz2002Introduction}
{Schmetz}, J., {Pili}, P., {Tjemkes}, S., {Just}, D., {Kerkmann}, J., {Rota},
  S., and {Ratier}, A. (2002).
\newblock {An Introduction to Meteosat Second Generation (MSG).}
\newblock {\em Bulletin of the American Meteorological Society}, 83:977--992.

\bibitem[Schn\"{o}rr, 1991]{Schnorr1991Determining}
Schn\"{o}rr, C. (1991).
\newblock Determining optical flow for irregular domains by minimizing
  quadratic functionals of a certain class.
\newblock {\em International Journal of Computer Vision}, 6(1):25--38.

\bibitem[Seemann et~al., 2008]{Seemann2008Development}
Seemann, S.~W., Borbas, E.~E., Knuteson, R.~O., Stephenson, G.~R., and Huang,
  H.-L. (2008).
\newblock Development of a global infrared land surface emissivity database for
  application to clear sky sounding retrievals from multispectral satellite
  radiance measurements.
\newblock {\em Journal of Applied Meteorology and Climatology}, 47(1):108--123.

\bibitem[Simoncelli et~al., 1991]{Simoncelli1991Probability}
Simoncelli, E., Adelson, E., and Heeger, D. (1991).
\newblock Probability distributions of optical flow.
\newblock In {\em Proceedings of the IEEE Computer Society Conference on
  Computer Vision and Pattern Recognition}, pages 310--315.

\bibitem[Simpson et~al., 2012]{Simpson2012In}
Simpson, D., Lindgren, F., and Rue, H. (2012).
\newblock In order to make spatial statistics computationally feasible, we need
  to forget about the covariance function.
\newblock {\em Environmetrics}, 23(1):65--74.

\bibitem[Xu et~al., 2005]{Xu2005Kernel-Based}
Xu, K., Wikle, C.~K., and Fox, N.~I. (2005).
\newblock A kernel-based spatio-temporal dynamical model for nowcasting weather
  radar reflectivities.
\newblock {\em Journal of the American Statistical Association},
  100(472):1133--1144.

\end{thebibliography}



\end{document}